\documentclass[fleqn,usenatbib]{mnras}

\usepackage{newtxtext,newtxmath}

\usepackage[T1]{fontenc}
\usepackage{ae,aecompl}

\usepackage{graphicx}	
\usepackage{amsmath}	
\usepackage{amssymb}	

\newcommand{\delp}{\frac{\partial}{\partial p}}
\newcommand{\delep}{\frac{\partial}{\partial \epsilon}}

\newcommand{\vect}[1]{\mbox{\boldmath${#1}$}}
\newcommand{\delf}[2]{\ensuremath{\frac{\partial #1}{\partial #2}}}
\newcommand{\df}[2]{\ensuremath{\frac{d #1}{d #2}}}

\newcommand{\msun}{M_{\odot}}

\title[High-energy particles in hot accretion flows]{Acceleration and escape processes of high-energy particles in turbulence inside hot accretion flows}

\author[S.S. Kimura, K. Tomida, K. Murase]{
Shigeo S. Kimura$^{1,2,3}$\thanks{E-mail: szk323@psu.edu},
Kengo Tomida$^{4,5}$,
Kohta Murase$^{1,2,3,6}$
\\
$^1$Department of Physics, Pennsylvania State University, University Park, Pennsylvania, 16802, USA\\
$^2$Center for Particle and Gravitational Astrophysics, Pennsylvania State University, University Park, Pennsylvania, 16802, USA\\
$^3$Department of Astronomy \& Astrophysics, Pennsylvania State University, University Park, Pennsylvania, 16802, USA\\
$^4$Department of Earth \& Space Science, Osaka University, Osaka, 560-0043, Japan\\
$^5$Department of Astrophysical Sciences, Princeton University, Princeton, NJ 08544, USA\\
$^6$Center for Gravitational Physics, Yukawa Institute for Theoretical Physics, Kyoto University, Kyoto 606-8502 Japan\\
}

\date{Accepted XXX. Received YYY; in original form ZZZ}

\pubyear{2018}

\begin{document}
\label{firstpage}
\pagerange{\pageref{firstpage}--\pageref{lastpage}}
\maketitle

 \begin{abstract}
  We investigate acceleration and propagation processes of high-energy particles inside hot accretion flows. The magnetorotational instability (MRI) creates turbulence inside accretion flows, which triggers magnetic reconnection and may produce non-thermal particles. They can be further accelerated stochastically by the turbulence. To probe the properties of such relativistic particles, we perform magnetohydrodynamic simulations to obtain the turbulent fields generated by the MRI, and calculate orbits of the high-energy particles using snapshot data of the MRI turbulence. We find that the particle acceleration is described by a diffusion phenomenon in energy space with a diffusion coefficient of the hard-sphere type: $D_\epsilon\propto \epsilon^2$, where $\epsilon$ is the particle energy. Eddies in the largest scale of the turbulence play a dominant role in the acceleration process. On the other hand, the stochastic behaviour in configuration space is not usual diffusion but superdiffusion: the radial displacement increases with time faster than that in the normal diffusion. Also, the magnetic field configuration in the hot accretion flow creates outward bulk motion of high-energy particles. This bulk motion is more effective than the diffusive motion for higher energy particles. Our results imply that typical active galactic nuclei that host hot accretion flows can accelerate CRs up to $\epsilon\sim 0.1-10$ PeV.
\end{abstract}


\section{Introduction}\label{sec:intro}

Mass accretion onto a compact object powers broadband emissions from active galactic nuclei (AGN) and Galactic X-ray binaries. Hot accretion flows are formed when an accretion rate is sufficiently lower than the Eddington accretion rate, $\dot M_{\rm Edd}= L_{\rm Edd}/c^2$ \citep{Ich77a,ny94}, which are believed to be realized in our Galactic center \citep[Sgr A*;][]{NYM95a,mmk97}, low-luminosity active galactic nuclei \citep[LLAGNs;][]{nem+06,nse14}, and X-ray binaries in the low-hard state \citep{EMN97a,ycn05}.
Plasma in hot accretion flows can be collisionless in the sense that the thermalization timescale is longer than the dynamical timescale \citep{tk85,mq97,qg99}, which motivates us to consider the existence of non-thermal particles and the resulting emissions \citep{OPN00a,tt12,ktt14,kmt15,CNS17a}.
Although the spectrum of Sgr A* can be fit by the hot accretion flow models only with thermal electrons \citep{NYM95a,mmk97}, the models with non-thermal component match the observations better \citep{yqn03,BOP16a}.
Non-thermal protons  (hereafter, we call them cosmic rays; CRs) are more commonly expected to exist in hot accretion flows. 
They interact with thermal protons and photons, leading to gamma-ray and neutrino production. 
Some predictions have been made using one-dimensional modeling \citep{mnk97,OM03a,nxs13}, and observations of nearby Seyferts and LLAGNs by the {\it Fermi} satellite give some hints and put limits on the CR production efficiency \citep{WNX15a,WN17a}. 

Most of the previous studies used a single power-law distribution for the non-thermal components, $dN/d\epsilon\propto \epsilon^{-s}$ with $2\lesssim s\lesssim 4$, where $\epsilon$ is the particle energy.
Such a power-law distribution is expected if CRs are produced by first-order Fermi mechanisms, such as the diffusive shock acceleration \citep{Bel78a,BO78a}. 
However, it is unclear whether such a single power-law distribution is achieved, because the accretion flows are unlikely to have a strong shock. Although shocked accretion flows may be formed in hot accretion flows \citep[e.g.,][]{LB05a,bdl08}, we do not observe such structures in the multi-dimensional global hydrodynamic simulations \citep{YN14a}.
In the accretion flows without shocks, CRs are expected to be produced by magnetic reconnection \citep[e.g.,][]{Hos12a} and/or stochastic acceleration by turbulence \citep[e.g.,][]{lyn+14}.
Inside accretion flows, magnetrotational instability (MRI) generates strong turbulence and induces magnetic reconnection \citep[e.g.,][]{bh91,BH98a,SI01a}.
Recent Particle-In-Cell (PIC) simulations show that when MRI takes place in collisionless plasma, magnetic reconnection produces non-thermal particles \citep{riq+12,hos13,hos15,KSQ16a}. 
These non-thermal particles can further be accelerated stochastically through interactions with larger scale eddies.
However, current PIC simulations cannot track such a late time phase because of the computational limitation, although recent developments of computational resources and techniques partially enable us to simulate particle acceleration in turbulence \citep{CS18a,ZUW18a}. 
The stochastic particle acceleration by magnetohydrodynamic (MHD) turbulence is often modeled as a diffusion phenomenon in energy space \citep[e.g.,][]{BE87a}, which has been applied to various astrophysical objects such as galaxy clusters \citep[e.g.,][]{Bla00a,BL07a,FAK16a}, gamma-ray bursts \citep[e.g.,][]{AT09a,MAT12a}, radio-lobes of radio galaxies \citep[e.g.][]{HCF09a,ort09}, and blazars \citep[e.g.,][]{KGM06a,ATK14a}.
Engaging this stochastic acceleration model to the hot accretion flow at the Galactic center, we can explain flares of Sgr A* \citep{LPM04a}, TeV gamma rays from the Galactic Center \citep{LMP06a,FKM15a}, and perhaps PeV CRs observed at the Earth \citep{FMK17a}. 
In addition, \citet{kmt15} showed that using the acceleration model, hot accretion flows in LLAGNs can reproduce the high-energy neutrinos detected by IceCube.
Note that the model leads to a very hard spectrum, $-1\le s \le 0$, compared to the shock acceleration.

In the stochastic acceleration model, the diffusion coefficient in energy space is approximated by a power-law function of energy, $D_\epsilon\approx D_0 (\epsilon/\epsilon_0)^q$.
The values of $q$ and $D_0$ depend on the power spectrum of the MHD turbulence and interaction processes between CRs and MHD waves \citep[e.g.,][]{CL06a}. 
For example, gyro resonant scattering by Alfven modes makes the value of $q$ equal to the slope of the power spectrum of the turbulence \citep[e.g.,][]{dml96,bld06,sp08}. 
The turbulent strength, $(\delta B/B_0)^2$, is related to $D_0$, and analytic theories used in the works above assume that the turbulent strength is smaller than unity. 
However, this condition is likely to be violated in weakly magnetized accretion flows according to MHD simulations \citep[e.g.,][]{SP01a,Mck06a,STK14a}. 
Applicability of the analytic models to the strong turbulence has been investigated using test particle simulations, but it is still controversial. 
The turbulence is usually provided by a superposition of plane waves in the Fourier space \citep[e.g.,][]{ort09,FM14a,TIN15a}, or driven by some algorithms \citep[e.g.,][]{DMS03a,TWJ14a,lyn+14}. 
These studies are useful to investigate features of the stochastic acceleration owing to their controllablity of the turbulence. However, each astronomical object has a different driving mechanism of turbulence, which may lead to a different behaviour of the CR particles (see \citealt{RII16a} for supernova remnants and \citealt{PVL16a} for pulsar wind nebulae). 

\citet{KTS16a} performed test-particle simulations in the MRI turbulence using the shearing box approximation \citep{HGB95a}. However, the shearing box approximation has a few inconsistencies with the hot accretion flows, such as geometrical thickness and non-negligible advection cooling \citep{ny94}. More importantly, escape of CRs cannot be implemented in a realistic manner. In this paper, we present results of global simulations, which enables us to investigate behaviours of the high-energy CRs more consistently. We perform MHD simulations to model hot and turbulent accretion flows, and solve orbits of test particles using the snapshot data of the MHD simulations. This paper is organized as follows. First, we describe the global MHD simulations dedicated to the hot accretion flows in Section \ref{sec:mhd}. Then, we show the results of the test-particle simulations in Section \ref{sec:particle}. We  discuss implications and future directions in Section \ref{sec:discussion}, and summarize our results in Section \ref{sec:summary}.%


\section{Properties of the MRI turbulence}\label{sec:mhd}

\subsection{Setup for MHD simulations}\label{sec:setup-mhd}

We use the Athena++ code\footnote{https://princetonuniversity.github.io/athena/} to solve the set of the ideal MHD equations \citep[][; Stone et al. in prep.]{SGT08a}: 
\begin{eqnarray}
 \delf{\rho}{T}+\vect\nabla \cdot \left(\rho\vect V \right)&=&0,\\
 \delf{(\rho\vect V)}{T} + \vect\nabla\cdot\left(\rho \vect V \vect V - {\vect B \vect B\over 4\pi} +P^*\mathbb I \right) &=& -\rho\vect\nabla \Phi ,\\
 \delf{E_{\rm tot}}{T} + \vect\nabla\cdot\left[\left(E_{\rm tot}+P^*\right) \vect V - {\vect B\cdot\vect V\over 4\pi}\vect B \right] &=& -\rho \vect V\cdot \vect\nabla\Phi,\\
 \delf{\vect B}{T} - \vect\nabla \times \left(\vect V \times \vect B\right)&=&0, 
\end{eqnarray}
where $T$ is the time for the MHD calculations, $\rho$ is the density, $\vect{V}$ is the velocity of the MHD fluid, $\vect B$ is the magnetic field, $P^*=P+B^2/(8\pi)$ is the total pressure, $P$ is the gas pressure, $\mathbb I$ is the unit tensor, and $\Phi$ is the gravitational potential. The total energy of the fluid is written as 
\begin{equation}
 E_{\rm tot}=E_{\rm th} + \frac 1 2 \rho V^2 + {B^2\over 8\pi},
\end{equation}
and we use the equation of state for ideal gas, $P=(\gamma_s-1)E_{\rm th}$ ($\gamma_s=5/3$ is the specific heat ratio and $E_{\rm th}$ is the thermal energy).
We solve the MHD equations in the spherical polar coordinate, $(R,~\theta,~\phi)$, using the second-order van-Leer integrator, the second-order piecewise linear reconstruction, the Harten-Lax-van Leer Discontinuities (HLLD) approximate Riemann solver \citep{MK05a}, and the constrained transport scheme. We use the Newtonian gravitational potential, $\Phi=-GM/R$, where $G$ is the gravitational constant and $M$ is the mass of the central black hole (BH). With this potential, we do not have to specify the values of the BH mass and the distance from the BH (see the last paragraph of this subsection about the unit for MHD calculations). Hence, we can use the same MHD data set to multiple particle simulations with various parameters by scaling the physical quantities (see Section \ref{sec:setup-particle}). On the other hand, with the pseudo-Newtonian potential commonly used in the literature \citep{PW80a}, we need to give specific values of both the mass of the BH and the radial distance from the BH to simulate the MHD turbulence. This requires more MHD data sets than those with the Newtonian potential, making it impossible to investigate wide parameter space due to a limited computational resource. 

The initial condition for the MHD simulations is an equilibrium torus with poloidal magnetic field loops embedded in a non-rotating gas of uniform density without magnetic fields as in \citet{SP01a} \citep[see also][]{PP84a,SPB99a}. 
Within the torus, the pressure is related to the density as $P=K\rho^{\gamma_s}$. The density distribution is represented as
\begin{equation}
 {P\over \rho}= {GM\over (n+1) R_c}\left[{R_c\over R}-\frac 1 2 \left({R_c\over R\sin\theta}\right)^2-{1\over 2d}\right],
\end{equation}
where $n=(\gamma_s-1)^{-1}$ is the polytropic index, $R_c$ is the radius of the density maximum of the torus, and $d$ is the distortion parameter. We set $\rho=\rho_{\rm env}$ and $P=\rho_{\rm env}/R$ outside the torus. We give the magnetic field through the vector potential: $A_\phi=\rho^2/\beta_0$ and $A_r=A_\theta=0$ in the torus and $\vect A=0$ otherwise. This vector potential produces magnetic fields parallel to the density contours. 

Our computational grids extend from $R_{\rm min}=0.1R_c$ to $R_{\rm max}=4.0R_c$, from $\theta_{\rm min}=\pi/6$ to $\theta_{\rm max}=5\pi/6$, and all the azimuthal angle (i.e. $0\le\phi\le2\pi$). Alfven velocity at the polar region is expected to become very high as the calculation proceeds \citep[e.g.,][]{Mck06a}, so the polar regions $\theta<\pi/6$ and $\theta>5\pi/6$ are truncated to reduce the computational time. We use the outflow boundary conditions for both inner and outer boundaries of the $R$ and $\theta$ directions. We performed some simulations with other values of $\theta_{\rm max}$ and $\theta_{\rm min}$, and confirmed that they do not affect the features of turbulence discussed in Section \ref{sec:result-mhd}.
For our fiducial run, the grid points for the $\theta$ and $\phi$ directions are distributed equally in linear space, while the grids in the radial direction are equally spaced logarithmically.

We perform the simulations with the unit of $R_c=1$, $GM=1$, and $\rho_c=1$, where $\rho_c$ is the maximum density at the torus. We set $\beta_0=10^3$ and $d=1.5$. For this value of $d$, the innermost radius of the initial torus is located at $R_{\rm trc}\simeq 0.63R_c$. 
We show the results of four runs; they are different only in resolutions and grid spacing as tabulated in Table \ref{tab:MHD}. For runs A, B, and C, the radial grids are equally spaced logarithmically, while for run D, they are equally distributed in the linear space.

\begin{table*}
\begin{center}
\caption{Parameters and physical quantities for MHD simulations. Here, $\langle X\rangle_{\mathcal{V}}$ indicates the volume-averaged values of $0.1\le R/R_c \le 0.6$.  \label{tab:MHD}}
\begin{tabular}{|c|cccc|}
\hline
run & A & B & C & D \\
\hline
$(N_r,~N_\theta,~N_\phi)$ for $R<R_{\rm max}$ & (640, 320, 768) & (384, 192, 512) &(256, 128, 384)& (960, 192, 512) \\
$(N_r,~N_\theta,~N_\phi)$ for $R<0.6R_c$ & (320, 320, 768) & (192, 192, 512) &(127, 128, 384)& (166, 192, 512) \\
$\Delta R_{i+1}/\Delta R_i$ & 1.006 & 1.01 & 1.015 & 1 \\
$\alpha_{\rm SS,M}$ & 0.022 &  0.025 & 0.023& 0.021 \\
$\alpha_{\rm mag}$ & 0.45 & 0.46 & 0.38  & 0.47\\
$\langle B_R^2\rangle_{\mathcal{V}}/\langle B_\phi^2\rangle_{\mathcal{V}}$ & 0.16 & 0.16 & 0.10 & 0.16 \\
\hline
$\langle B^2\rangle_{\mathcal{V}}/ \langle8\pi P\rangle_{\mathcal{V}}$ & 0.050  & 0.053 & 0.061 & 0.044  \\
$\langle B_R^2\rangle_{\mathcal{V}}/\langle8\pi P\rangle_{\mathcal{V}}$ & $6.7\times10^{-3}$ & $7.0\times10^{-3}$ & $5.5\times10^{-3}$ & $5.7\times10^{-3}$\\
$\langle B_\theta^2\rangle_{\mathcal{V}}/\langle8\pi P\rangle_{\mathcal{V}}$ & $2.8\times10^{-3}$& $2.9\times10^{-3}$ & $2.0\times10^{-3}$ & $2.2\times10^{-3}$ \\
$\langle B_\phi^2\rangle_{\mathcal{V}}/\langle8\pi P\rangle_{\mathcal{V}}$ & 0.040 & 0.043 & 0.053& 0.036 \\
$\langle cE_R^2\rangle_{\mathcal{V}}/\langle8\pi P\rangle_{\mathcal{V}}$ & $6.9\times10^{-3}$& $6.7\times10^{-3}$ & $4.7\times10^{-3}$ &$5.0\times10^{-3}$ \\ 
$\langle cE_\theta^2\rangle_{\mathcal{V}}/\langle8\pi P\rangle_{\mathcal{V}}$ & 0.015 & 0.015 & 0.011 &0.012 \\
$\langle cE_\phi^2\rangle_{\mathcal{V}}/\langle8\pi P\rangle_{\mathcal{V}}$ & $5.6\times10^{-5}$  & $7.4\times10^{-5}$ & $6.8\times10^{-5}$& $5.5\times10^{-5}$\\ 
\hline
\end{tabular}
\end{center}
\end{table*}

\subsection{Results of MHD simulations}\label{sec:result-mhd}

We are interested in features of the turbulence inside the disc rather than the lower density corona located above the disc. Thus, we calculate mass-weighted average values of physical quantities. We use the snapshot data at $T\Omega_c=20\pi$, where $\Omega_c$ is the Kepler angular velocity at $R_c$, unless otherwise noted. We also analyzed the snapshot data at $T\Omega_c=18\pi$ and confirmed that the basic features are the same.

First, we compute volume-averaged quantities of the MRI turbulence:
\begin{equation}
 \langle X \rangle_{\mathcal{V}} = {\int \rho X R^2 \sin\theta dRd\theta d\phi\over \int \rho R^2 \sin\theta dRd\theta d\phi}.
\end{equation}
The radial integration is performed from $R=R_{\rm min}$ to $R=0.6R_c$, where the outer radius is chosen so as to be smaller than the innermost radius of the initial torus. 
The integration for the other two directions are performed over all the computational region.
The grid numbers within the integration region are tabulated in Table \ref{tab:MHD}.

   \begin{figure}
   \begin{center}
    \includegraphics[width=\linewidth]{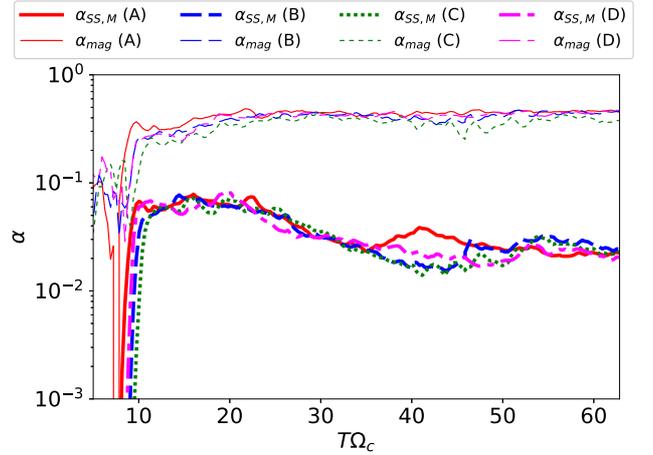}
    \caption{Time evolutions of $\alpha_{\rm SS,M}$ (thick lines) and $\alpha_{\rm mag}$ (thin lines). The solid, dashed, dotted, and dot-dashed lines are for runs A, B, C, and D, respectively. We can see that these lines converge for $T\Omega_c\gtrsim 50$, which indicates realization of the quasi-steady states. }
    \label{fig:alpha}
   \end{center}
   \end{figure}

  \begin{figure}
   \begin{center}
    \includegraphics[width=\linewidth]{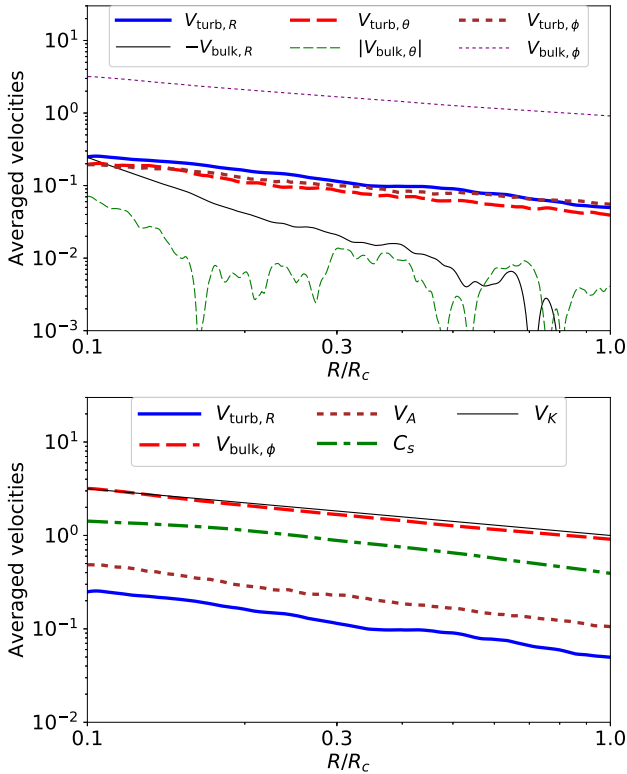}
    \caption{Radial distributions of the velocities for run A. Upper panel: Comparison of turbulent velocities and bulk velocities for $R$, $\theta$, and $\phi$ components. Lower panel: Comparison of turbulent, rotation, sound, and Alfven velocities. The Keplerian velocity is also shown as the thin solid line.}
    \label{fig:velocity}
   \end{center}
  \end{figure}

  \begin{figure*}
   \begin{minipage}{0.33\hsize}
   \begin{center}
    \includegraphics[width=\linewidth]{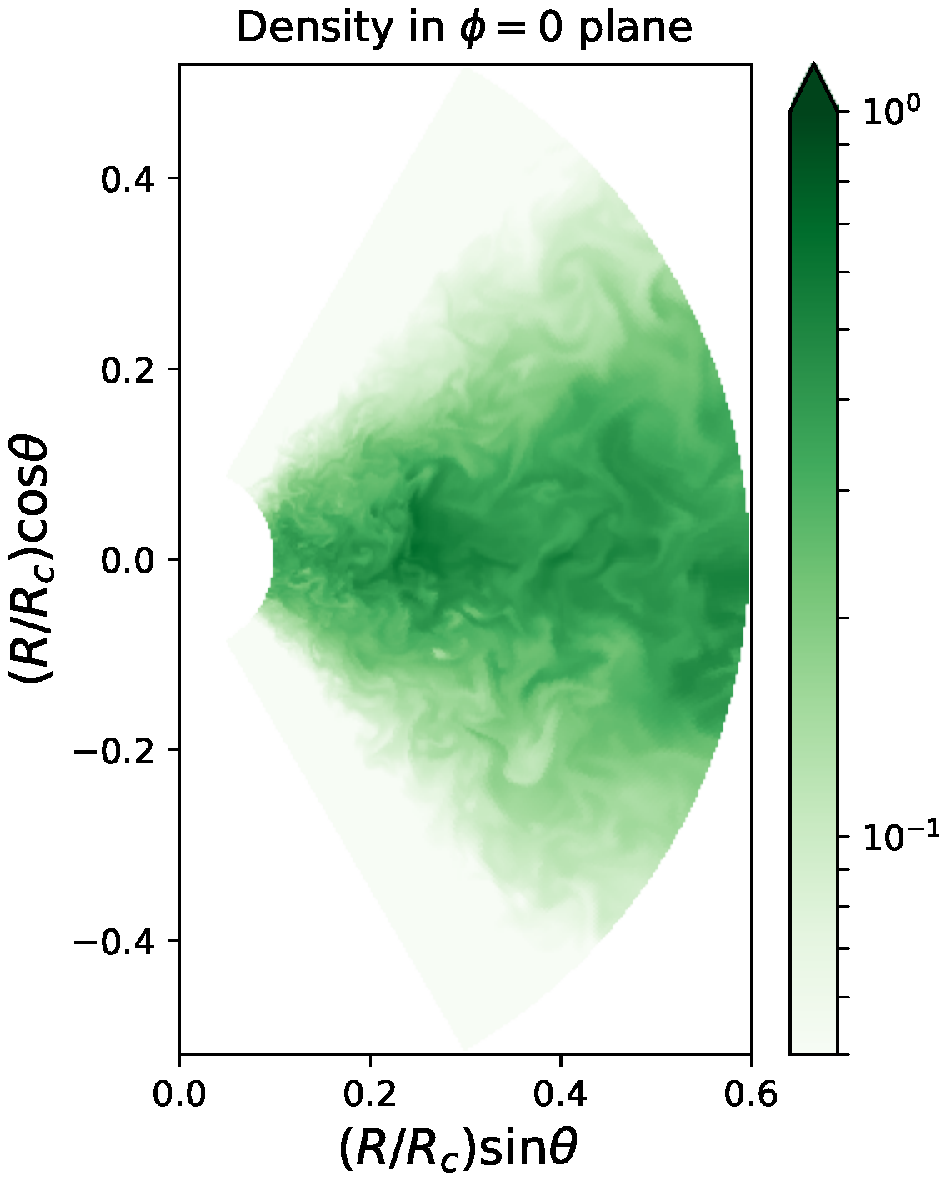}
   \end{center}
   \end{minipage}
   \begin{minipage}{0.33\hsize}
   \begin{center}
    \includegraphics[width=\linewidth]{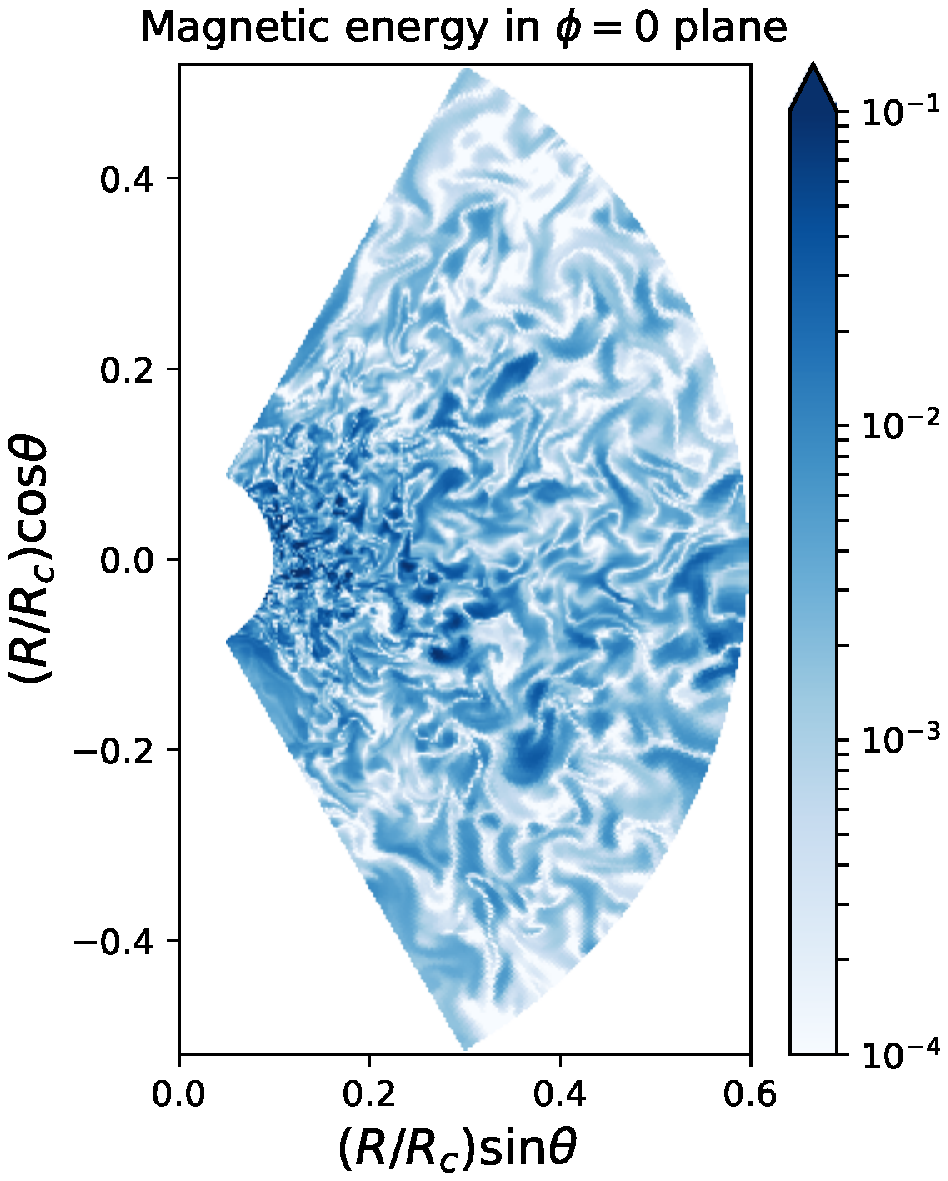}
   \end{center}
   \end{minipage}
   \begin{minipage}{0.33\hsize}
   \begin{center}
    \includegraphics[width=\linewidth]{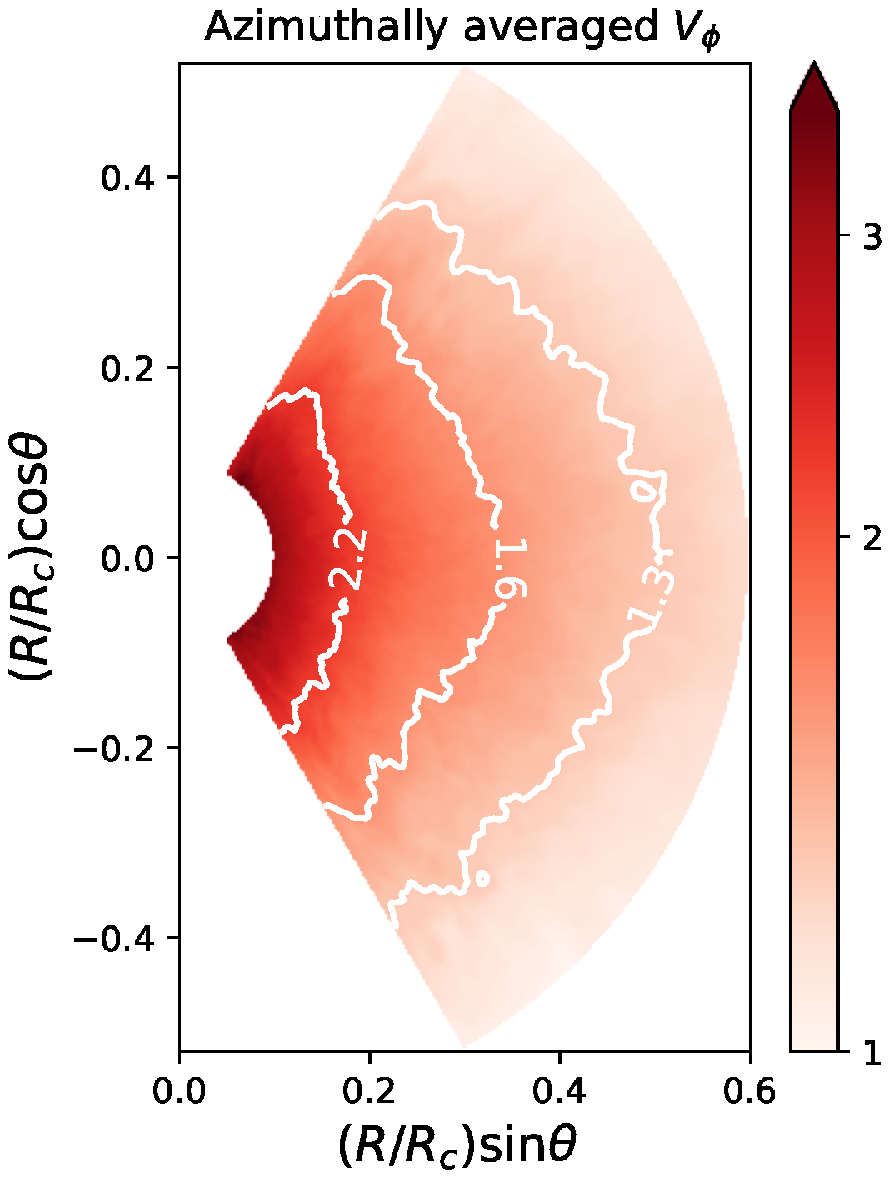}
   \end{center}
   \end{minipage}
    \caption{Colormaps in the meridional plane for run A. Left: density on the $\phi=0$ plane. Center: magnetic energy density, $B^2/(8\pi)$, on the $\phi=0$ plane. Right: Azimuthally averaged $V_\phi$, $\langle V_\phi \rangle_{\mathcal{L}}$,  on the $R-\phi$ plane. The white lines are iso-contours of  $\langle V_\phi \rangle_{\mathcal{L}}$. }
    \label{fig:r-theta}
  \end{figure*}

  \begin{figure}
   \begin{center}
    \includegraphics[width=\linewidth]{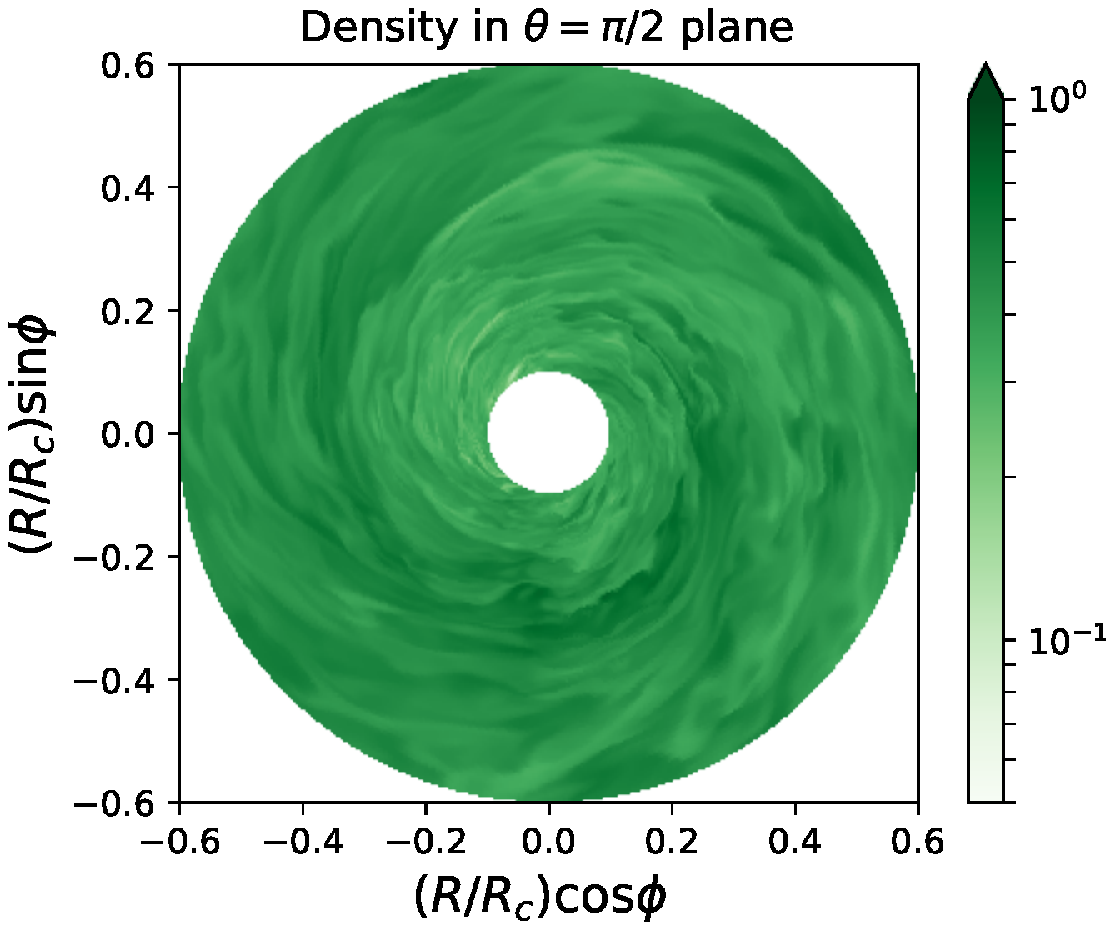}
    \includegraphics[width=\linewidth]{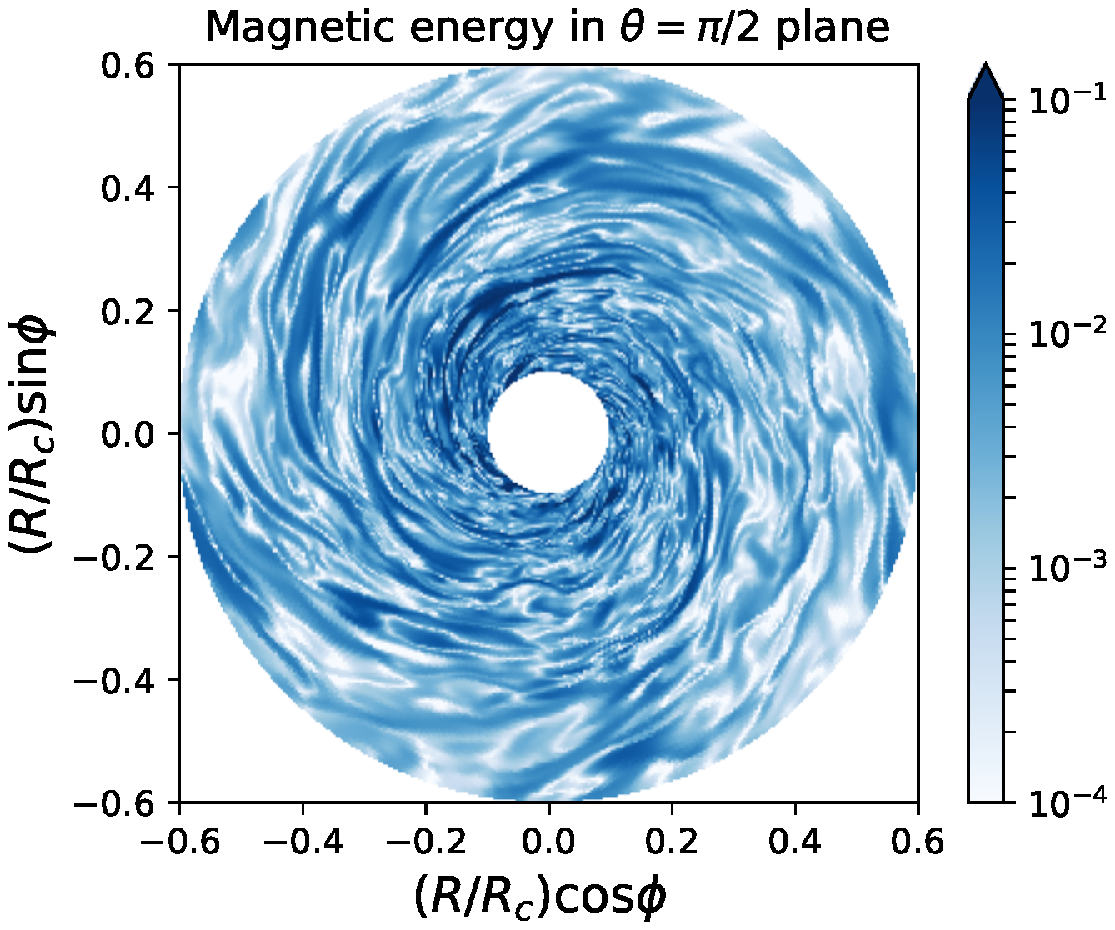}
    \caption{Colormaps in the equatorial plane for run A. The upper and lower panels show the density and the magnetic energy density, respectively.}
    \label{fig:r-phi}
   \end{center}
  \end{figure}

  \begin{figure*}
   \begin{center}
    \includegraphics[width=\linewidth]{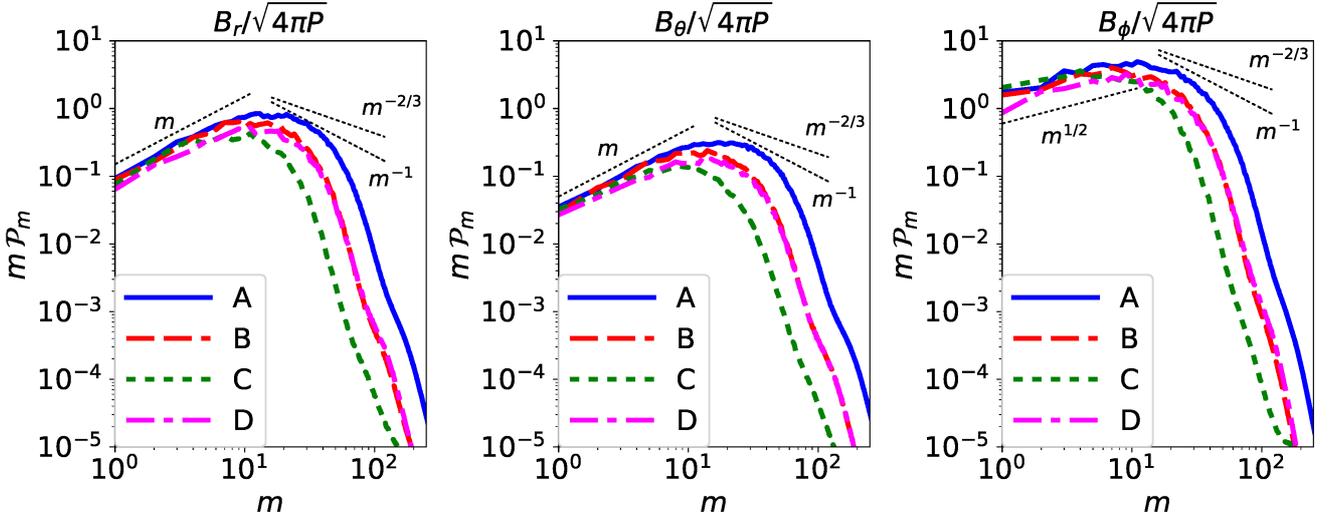}
    \caption{Power spectra for the magnetic fields in the azimuthal direction. The left, middle, and right panels show the power spectra for $B_r$, $B_\theta$, and $B_\phi$, respectively. The solid, dashed, dotted, and dot-dashed lines are for runs A, B, C, and D, respectively.}
    \label{fig:pow}
   \end{center}
  \end{figure*}

Figure \ref{fig:alpha} shows the time evolution of Maxwell stresses normalized by the gas pressure, $\alpha_{\rm SS,M}$, and by the magnetic pressure, $\alpha_{\rm mag}$. They are defined as 
\begin{eqnarray}
 \alpha_{\rm SS,M}= {\langle B_r B_\phi\rangle_{\mathcal{V}} \over\langle 4\pi P\rangle_{\mathcal{V}}}, \\
 \alpha_{\rm mag}= {\langle 2B_r B_\phi\rangle_{\mathcal{V}} \over \langle B^2\rangle_{\mathcal{V}}} 
\end{eqnarray}
respectively. The subscript SS represents \citet{ss73} in which the $\alpha$ parameter is introduced for the first time\footnote{The definition of the $\alpha$ parameter in \citet{ss73} includes the Reynolds stress, but we do not include it to $\alpha_{\rm SS,M}$ because it is sub-dominant \citep[e.g.,][]{SI14a}.}. We can see that these values converge at $\alpha_{\rm SS,M}\simeq 0.02-0.03$ and $\alpha_{\rm mag}\simeq0.4-0.5$ for $T\Omega_c \gtrsim 50$. This displays that the quasi-steady state is realized. We compute the mass accretion rates and find that they are roughly constant in both time and radius, which also indicates the realization of the quasi-steady state.
$\alpha_{\rm mag}$ is used as an indicator of the numerical convergence of the MRI turbulence \citep{HGK11a,HRG13a}. We tabulate the values of $\alpha_{\rm mag}$ at the end of the calculations. Sufficiently high-resolution simulations give $\alpha_{\rm mag}\simeq0.45$, which is seen in runs A, B, and D. The ratio of the radial magnetic energy to the azimuthal one, $\langle B_R^2\rangle_{\mathcal{V}}/\langle B_\phi^2\rangle_{\mathcal{V}}$ is also useful in diagnosis \citep{HGK11a,HRG13a}. The values of the ratio converge to 0.15--0.16 for runs A, B and D as tabulated in Table \ref{tab:MHD}. Hence, the grid numbers for our simulations are high enough to follow the features of the MRI turbulence except for run C.


The inverse of plasma beta, $B^2/(8\pi P)$, and each component of magnetic and electric fields are also tabulated in Table \ref{tab:MHD}. Although the MRI amplifies the magnetic field, the gas pressure is still stronger than the magnetic pressure. In the accretion flow, $B_\phi$ dominates over the other two components because the shear motion stretches the magnetic field. The electric field is computed by the ideal MHD condition, 
\begin{equation}
 \vect E = - {\vect{V}\times\vect B\over c}.\label{eq:MHDcondition}
\end{equation}
Since $B_\phi$ and $V_\phi$ are dominant, $E_r$ and $E_\theta$ are stronger than $E_\phi$. 
Note that the electric field in the accretion flow is much weaker than the magnetic field, since $V/c <1$. 



The radial profiles of the velocities are estimated to be 
\begin{equation}
\langle X \rangle_{\mathcal{S}}(R)={\int \rho X \sin\theta d\theta d\phi \over \int \rho \sin\theta d\theta d\phi}.
\end{equation}
The fluid velocity strongly affects the orbits of the test particles, since the drift velocities of the test particles are the same as the fluid velocity perpendicular to the magnetic field.
Each component of the fluid velocity is divided into two parts, $V_i=V_{{\rm bul},i}+V_{{\rm tur},i}$. For $V_R$ and $V_\phi$, assuming that $\langle V_{{\rm tur},i}\rangle_{\mathcal{S}}=0$, we obtain $V_{{\rm bul},i}=\langle V_i\rangle_{\mathcal{S}}$. For $V_\theta$, we define $V_{{\rm bul},\theta}=\langle V_\theta {\rm sgn}(\cos\theta) \rangle_{\mathcal{S}}$. Then, the turbulent component for each direction is estimated to be $V_{{\rm tur},i}=\sqrt{\langle V_i^2\rangle_{\mathcal{S}}-V_{{\rm bul},i}^2}$.
We plot $V_{{\rm bul},i}$ and $V_{{\rm tur},i}$ in the upper panel of Figure \ref{fig:velocity}. The turbulent components dominate over the bulk components for $V_R$ and  $V_\theta$. On the other hand, $V_{{\rm bul},\phi}$ is much higher than its turbulent component. The turbulent components are comparable each other, but $V_{{\rm tur},R}$ is slightly higher than the other components. 

The lower panel of Figure \ref{fig:velocity} shows comparison of the turbulent velocity ($V_{{\rm tur},R}$), rotation velocity ($V_{{\rm bul},\phi}$), sound velocity ($C_s=\sqrt{\gamma_s P/\rho}$), and Alfven velocity ($V_A=B/\sqrt{4\pi \rho}$).
We see that the rotation velocity is the fastest of the four and its value is similar to the Keplerian velocity ($V_K=\sqrt{GM/R}$) shown as the thin line. The sound velocity is the second fastest, and Alfven velocity follows it.  This means that the accretion flow consists of high-$\beta$ plasma and that the turbulence is sub-sonic and sub-Alfvenic.
These results are consistent with the previous simulations \citep[e.g.,][]{SP01a,MM03a}. 

Next, we briefly discuss two-dimensional structures. 
The left and middle panels of Figure \ref{fig:r-theta} show the colormaps of the density and the magnetic energy on the $\phi=0$ plane, respectively.
We can see from the left panel that the accretion flow is geometrically thick with the aspect ratio $H/R\sim 0.5$, and the density at midplane is much higher than that around the polar boundaries. The middle panel shows that the magnetic field is strongly turbulent not only in the disc region ($|\cos\theta| \lesssim 0.45$) but also in the corona region  ($|\cos\theta| \gtrsim 0.45$).
The right panel  represents contours of the azimuthally averaged $V_\phi$:
\begin{equation}
 \langle V_\phi \rangle_{\mathcal{L}}(R,~\phi)={\int \rho V_\phi\phi \over\int \rho d\phi }.
\end{equation}
We can see that the iso-$V_\phi$ surface depends on the spherical radius, $R$, rather than the cylindrical radius, $R\sin\theta$. 
This allows us to use $V_{\rm{bul},\phi}$ as the background velocity for analyses of the test-particle simulations in Section \ref{sec:result-particle}.

Figure \ref{fig:r-phi} plots the colormaps of the density (upper) and the magnetic energy (lower) on the equatorial plane. The magnetic fields are frozen in the differentially rotating fluid elements that fall to the BH. This creates the spiral structure as seen in the figure. We can also see the fluctuation of the density is much smaller than that of the magnetic field energy density. This implies that the fast modes are a sub-dominant component in the MRI turbulence.

To clarify the importance of the modes of the MHD waves (fast, slow, and Alfven), we evaluate the Pearson correlation coefficients between the fluctuations of the density, $\delta\rho(R,~\theta,~\phi)=\rho-\langle \rho \rangle_{\mathcal{L}}$, and the magnetic energy, $\delta B^2(R,~\theta,~\phi)=B^2-\langle B^2 \rangle_{\mathcal{L}}$. According to the linear MHD wave theory, the fast mode has a positive correlation, the slow mode has a negative correlation, and the Alfven mode has no correlation. We evaluate the correlation coefficients as a function of $R$ and $\theta$, and average over them with weights associated with the area in the meridional plane. The resulting coefficients indicate that the density and magnetic energy are weakly anti-correlated: the value of the coefficient is $-0.22$ in the disc region ($|\cos\theta\lesssim 0.45|$) for run A. The lower resolution runs have higher coefficients, i.e., the anti-correlations are weaker, but no run has a positive correlation. Therefore, the fast modes do not play an important role in this system. This result is natural in the sub-Alfvenic and sub-sonic turbulence.


Finally, we discuss the azimuthal power spectra of the turbulence (cf. \citealt{SRS12a,SI14a}; see \citealt{PB13a} for three-dimensional power spectra).
We take the Fourier transformation in the azimuthal direction,
\begin{equation}
 X_m={1\over \sqrt{2\pi}}\int X \exp(-im\phi) d\phi,
\end{equation}
where $m=k_\phi R$ ($k_\phi$ is the wavenumber in the $\phi$ direction).
Then, we take the average of the power spectrum over the disc region:
\begin{equation}
\mathcal P_m={\int |X_m|^2 R  dR d\theta\over \int  R d R d\theta },
\end{equation}
where the integration region is set to be $0.1 R_c \le R \le 0.6 R_c$ and $|\cos\theta| \le 0.45$. 
We plot the power spectra, $m\mathcal P_m$, for the magnetic field in Figure \ref{fig:pow}. We can see that all the data sets have similar values for a larger scale of $m\lesssim 10$. The spectra for $B_r$ and $B_\theta$ are $m\mathcal P_m\propto m$, while those for $B_\phi$  are roughly $m\mathcal P_m\propto m^{1/2}$. For a smaller scale of $m\gtrsim N_\phi/10$, the spectra decrease with $m$ very rapidly for all the data sets because of the numerical dissipation. The power spectra peak at intermediate scale of $m\sim10-20$, depending on the resolution and component. These features are consistent with the previous calculations \citep{SRS12a,SI14a}.

The fastest growing mode of the MRI is approximated to be $L_{\rm MRI}\sim 2\pi V_A/\Omega$, where $\Omega$ is the angular velocity. Saturation of MRI turbulence is expected to be controlled either by the large-scale magnetic reconnection \citep{SI01a,san+04} or by the growth of the parasitic instabilities of Kelvin-Helmholtz modes \citep{GX94a,Pes10a}. These phenomena occur inside the disc, where the largest scale is the scale height, $H\approx C_s/\Omega$. Hence, the characteristic scale of the saturated MRI turbulence should be the smaller one of the two, $L_{\rm tur}\approx {\rm min} (L_{\rm MRI},~H)$.
From Figure \ref{fig:velocity}, we roughly see $V_A\sim V_{\phi,\rm bulk}/7$ and $C_s\sim V_{\phi,\rm bulk}/2$, leading to $L_{\rm MRI}\approx 2\pi R/7 > H\approx R/2$. Hence, $L_{\rm tur}=H\approx R/2$. This scale corresponds to $m\sim 13$, which is consistent with the peaks of the power spectra.

For the intermediate scale, we narrowly see that the spectra gradually decrease with $m$. Theoretically, fully developed Alfven turbulence results in $P_k \propto k_\perp^{-5/3}$ and $P_k\propto k_\parallel^{-2}$, where $P_k$ is the power spectrum, $k_\perp$ and $k_\parallel$ are the perpendicular and parallel wave numbers to the background magnetic field, respectively \citep{gs95}. 
Such an anisotropic cascade takes place with respect to the local magnetic field. In strong turbulence where the large-scale magnetic field is significantly tilted, the direction of the local magnetic field is not aligned. Then, the global Fourier analysis would smear out the local anisotropy, resulting in $P_k\propto k^{-5/3}$ in all the directions \citep{2000ApJ...539..273C}. 
However, we cannot clearly see the power-law shape in the power spectra of our simulations, due to the insufficient dynamic range. Simulations with a higher resolution and a higher-order reconstruction scheme are necessary to determine the power-law index in the inertial range. To observe the anisotropic feature, even more dedicated analyses reconstructing coordinates based on the local magnetic field will also be required.
Note that the shape of these power spectra is independent of the integration range of $R$, because the turbulence is generated by the same mechanism at all the radii.

\section{Behaviours of High-Energy Particles}\label{sec:particle}

\subsection{Setup for particle simulations}\label{sec:setup-particle}

We calculate orbits of relativistic particles to investigate behaviour of high-energy particles in the accretion flows. We ignore CR injection mechanisms because they are related to small-scale plasma processes. They should be investigated by other methods, such as PIC simulations \citep{hos15,KSQ16a}, which is beyond the scope of this paper.

We solve the relativistic equation of motion for each CR particle:
\begin{equation}
 {d \vect p \over dt} = e \left(\vect E+{ \vect{v} \times\vect B \over c} \right),\label{eq:eom}
\end{equation}
where $t$ is the time for particle calculation, $c$ is the speed of light, and $\vect p=\gamma m_p \vect{v}$, $\vect{v}$, $e$, $m_p$, and $\gamma=\sqrt{1-(v/c)^2}$ are the momentum, velocity, charge, mass, and Lorentz factor of the CR particle, respectively. Here, we neglect the gravity acting on the CR particle, since it is typically weaker than the electromagnetic force by more than ten orders of magnitude.
This equation is integrated using the Boris method \citep[e.g.,][]{BL91a}, which is often used in PIC simulations. In the particle simulations, we use $m_p$ and $e$ for protons, but we can scale our simulation results to the heavy nuclei using the rigidity $\mathcal R=\epsilon/Z$.

The snapshot data of the MHD simulations shown in Section \ref{sec:result-mhd} are used to obtain $\vect E$ and $\vect B$.
Since the MHD data contain the values of $\vect V$ and $\vect B$ only at the discrete grid points, we first interpolate $\vect B$ and $\vect V$ at the position of the particle using quadratic functions\footnote{Although we use the quadratic functions for the interpolation, the results are very similar if we use the linear interpolation.}. 
Then, we compute $\vect E$ through Equation (\ref{eq:MHDcondition}) using the interpolated $\vect B$ and $\vect V$.
This procedure guarantees $\vect E \cdot \vect B=0$, so artificial acceleration due to the interpolation is avoided. 

We initially distribute particles on a ring of $R=R_{\rm ini}$ and $\theta=\pi/2$. 
The energy distribution of the initial particles is monoenergetic and isotropic in the fluid frame (see Section \ref{sec:result-particle}  for the definition of the fluid frame).
The initial radius is fixed at $R_{\rm ini}=0.3R_c$ for simplicity. We performed the simulations with $R_{\rm ini}=0.2 R_c$, and checked that the results are almost unchanged.
The initial energy of the particle, $\epsilon_{\rm ini}$, is given so that the Larmor radius of the particle is equal to $\lambda_{\rm ini}$ times the grid scale: $r_L= \epsilon_{\rm ini}/(e c B_{\rm ave}) = \lambda_{\rm ini}\Delta x_{\rm ini}$, where $\Delta x_{\rm ini}={\rm min}(\Delta R_{\rm ini},~R_{\rm ini}\Delta \theta,~R_{\rm ini}\Delta \phi)$ is the grid scale at the initial ring.
The timestep of the particle calculation is determined by $\Delta t={\rm min}(\Delta t_L,\Delta t_x)$, where $\Delta t_L = C_{\rm safe} t_{L,\rm min} = 2\pi C_{\rm safe} \epsilon_{\rm ini}/(e c B_{\rm max})$ and $\Delta t_x = C_{\rm safe}\Delta x_{\rm min}/c$. Here, $B_{\rm max}$ is the maximum value of the magnetic field, $\Delta x_{\rm min}$ is the minimum length between the grids in the computational region, and $C_{\rm safe}$ represents the safety factor that determines the timestep. We set $C_{\rm safe}=0.01$. We performed some simulations with $C_{\rm safe}=0.001$, and confirmed that the results are unchanged by the values of $C_{\rm safe}$. As a fiducial value, we set $\lambda_{\rm ini}=4$. With a smaller value of $\lambda_{\rm ini}$, we cannot trace the resonant scattering process, while the particles escape from the computational region too quickly with a higher value of $\lambda_{\rm ini}$. 

The computational region for the particle simulations is the same with the MHD simulations except for the outer boundary in the $R$ direction. Since the dynamical structures of the outer parts of the MHD simulations are affected by the initial conditions, we set the outer boundary of the particle simulations to $R_{\rm esc}=0.6 R_c$. The particles that go beyond the computational region are removed from the simulation, and we stop the calculation when half of the particles escape from the computational region.

We solve the equations of motion for $N_p=2^{14}=16384$ particles using the MHD data sets shown in the previous section.
To solve the equation of motion, we need to convert the units used in the MHD calculations to those of our interest. 
The units of the mass, length, and time for the MHD calculations are written as $\mathcal L_u=R_c$, $\mathcal M_u=\rho_cR_c^3$, and $\mathcal T_u=\sqrt{R_c^3/(GM)}$, respectively. 
For our particle simulations, we rescale these units as
\begin{eqnarray}
\mathcal L_u=\chi R_s,\\
\mathcal T_u=\sqrt{{\mathcal L_u^3\over GM}},\\
\mathcal M_u=\eta \dot M_{\rm Edd} \mathcal T_u,
\end{eqnarray}
where $R_s=2GM/c^2$ is the Schwarzschild radius, $\dot M_{\rm Edd}=L_{\rm Edd}/c^2$ is the Eddington mass accretion rate ($L_{\rm Edd}$ is the Eddington luminosity), and $\chi$ and $\eta$ are the scaling factors of the length and the mass, respectively. The relation between $\eta$ and density is $\rho_c=\eta \dot M_{\rm Edd} \mathcal T_u/\mathcal L_u^3$, so a higher $\eta$ leads to a higher density. 

We choose the reference parameter set for the particle simulations so as to be consistent with our assumptions: hot accretion flows in LLAGNs with Newtonian gravity.
In our MHD simulations, mass accretion rate is written as  $\dot M\sim \dot m_{\rm sim} \mathcal M_u \mathcal T_u^{-1}$, where $\dot m_{\rm sim}\sim10^{-3}-10^{-2}$ is the resulting mass accretion rate in the MHD simulations. Then, the rescaled mass accretion rate is represented as $\dot M = \eta \dot m_{\rm sim}\dot M_{\rm Edd}$. For $\eta\lesssim 10$, this mass accretion rate is in the hot accretion flow regime, i.e.,  $\dot M\lesssim0.1 \dot M_{\rm Edd}$ \citep{ny95,xy12}.
The scale factor for the length, $\chi$, should be large enough to be consistent with the Newtonian gravity. For $\chi\ge20$, the initial radius, $R_{\rm ini}=0.3R_c$, is larger than $6R_s=2R_{\rm ISCO}$, where $R_{\rm ISCO}=3R_s$ is the innermost circular stable orbit (ISCO) for the Schwartzchild BH. 
Based on the considerations above, we set the reference parameters to $\chi=50$, $\eta=1$, and $M=10^8M_{\odot}$, which corresponds to typical low-luminosity AGNs, such as Seyferts or low-ionization nuclear emission-line regions (LINERs). This parameter set leads to 
\begin{eqnarray}
\mathcal L_u\simeq1.5\times10^{15} M_8 \chi_{1.7}\rm~cm  \\
\mathcal T_u\simeq4.9\times10^5  M_8 \chi_{1.7}^{3/2} \rm~s,\\
\mathcal M_u\simeq 6.9\times10^{30}M_8^2 \chi_{1.7}^{3/2}\eta_0  \rm~g,
\end{eqnarray}
where we use the notation $Q_x=10^x$ (unit for $M$ is M$_\odot$). 
The speed of light is $c\simeq10\chi_{1.7}^{1/2} \mathcal L_u \mathcal T_u^{-1}$ in this unit system.
We use the MHD data set of run A with $T\Omega_c=20\pi$ unless otherwise noted. The Larmor radius and timescale are $r_L\simeq 1.0\times10^{13} M_8 \chi_{1.7} \lambda_{\rm ini,0.6}$ cm and $t_L\simeq 2.1\times10^{3} M_8 \chi_{1.7} \lambda_{\rm ini,0.6}$ s, respectively.

\subsection{Results of particle simulations}\label{sec:result-particle}

\subsubsection{Orbits and momentum distribution}\label{sec:orbits}

  \begin{figure}
   \begin{center}
    \includegraphics[width=\linewidth]{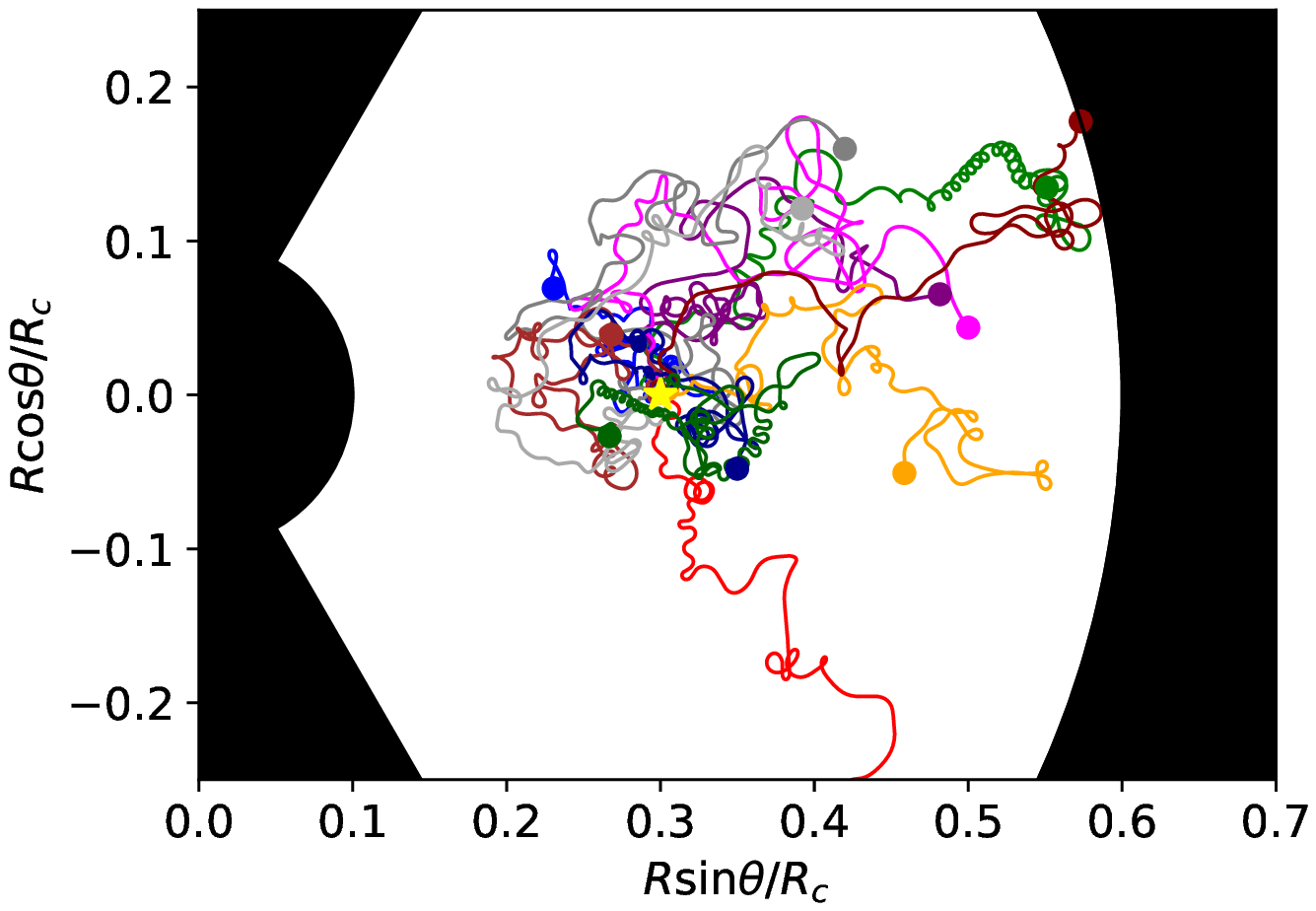}
    \includegraphics[width=\linewidth]{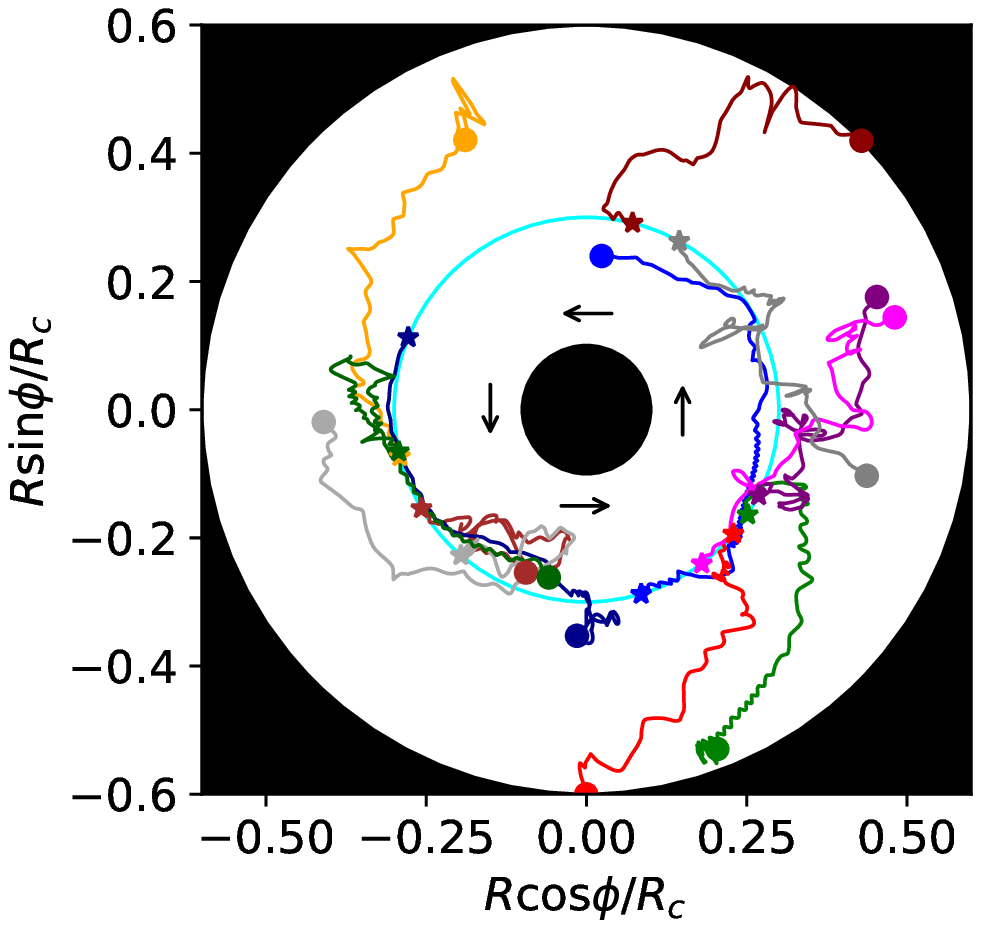}
    \caption{Orbits of test particles projected to the $R-\theta$ plane (upper panel) and the $R-\phi$ plane (lower panel) for $\lambda_{\rm ini}=4$. The initial and final position of the particles are shown by the stars and circles, respectively. In the bottom panel, the cyan circle and black arrows indicate the initial ring $R=R_{\rm ini}$ and the rotation direction, respectively. }
    \label{fig:orb}
   \end{center}
  \end{figure}

  \begin{figure}
   \begin{center}
    \includegraphics[width=\linewidth]{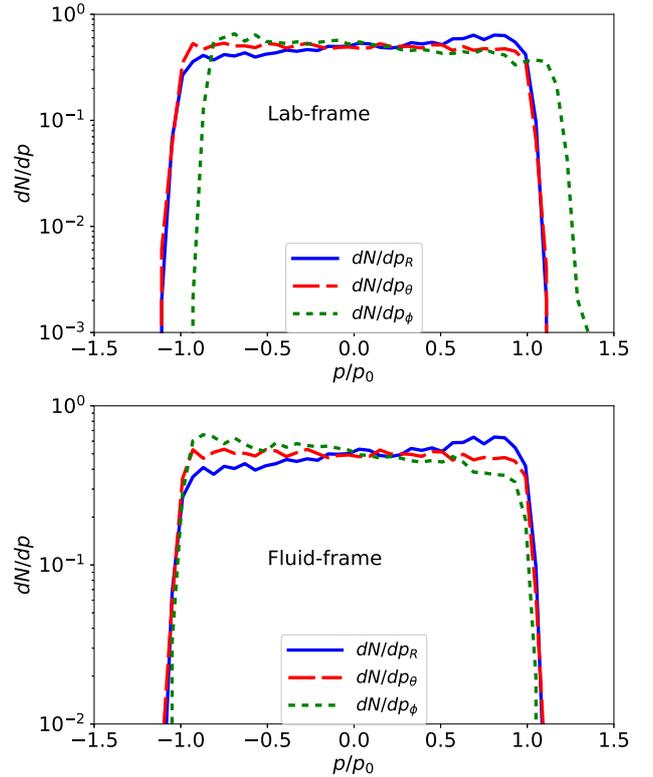}
    \caption{Momentum distributions at $t=10t_L$ in the lab frame (upper) and the fluid flame (lower) for $\lambda_{\rm ini}=4$. We can see a bulk motion in the lab-frame, while the bulk motion is not seen in the fluid frame.}
    \label{fig:pdist}
   \end{center}
  \end{figure}

The upper and lower panels of Figure \ref{fig:orb} show orbits of the test particles projected in the $R-\theta$ and $R-\phi$ planes, respectively. The particles spread in all the directions, but majority of the particles move outward in the $R$ direction rather than fall onto the BH or escape to the vertical direction.  The particles are likely to migrate to the direction at which the magnetic field is weak, partially due to the magnetic mirror force. The strong magnetic fields at the inner region prevent the particles from falling to the BH. Also, the magnetic fields at the high latitudes are not week, compare to those at the outer region (see Figure \ref{fig:r-theta}). At the end of the simulation, 68\% of all the escaping particles go out through the radial boundary with $|\cos\theta| \le 0.5$ for the case with $\lambda_{\rm ini}=4$.  This fraction is higher for lower $\lambda_{\rm ini}$, and vice versa for higher $\lambda_{\rm ini}$.  Higher energy particles can cross the magnetic field more easily, which may enhance the vertical diffusion.

From the lower panel, we find that the particles mostly travel along the $\phi$ direction, because the magnetic field is also directed to the $\phi$ direction. Interestingly, the outward-going particles tend to move the opposite direction to the background fluid. This arises from the magnetic field configuration. The accretion flow creates the spiral-shape magnetic fields as seen in Figure \ref{fig:r-phi}. When the CR particles stream along the field line outward, they counter-rotate with respect to the accretion flow.

The upper panel of Figure \ref{fig:pdist} shows the momentum distribution in each direction, $dN/dp_i$ ($i=r,~\theta,~\phi$)  measured in the lab frame at time $t=10 t_L$, where $t_L=2\pi \epsilon_{\rm ini}/(e c B_{\rm ini})$ and $B_{\rm ini}=\sqrt{\langle B^2\rangle_{\mathcal{S}}}$ at $R=R_{\rm ini}$. The momentum distribution is anisotropic: there is a bulk rotational motion.  This is because  the background fluid motion creates the electric field that induces $\vect E\times\vect B$ drift.

We compute the momentum distribution in the fluid frame by performing Lorentz transformation of the particle momenta. Since $V_\phi$ dominates over the other component, we approximate the background velocity to be 
\begin{equation}
\vect V_{\rm bg}=V_{{\rm bul},\phi}\vect e_\phi, \label{eq:vbg}
\end{equation}
where $\vect e_\phi$ is the unit vector of the $\phi$ direction and $V_{{\rm bul},\phi}$ is independent of $\theta$. The bottom panel shows the momentum distribution in the fluid frame, where we can see no bulk rotational motion. In the following sections, we use the energy distribution in the fluid frame. Note that the particle distribution is slightly anisotropic: the particles tend to have positive $p_R$ and negative $p_\phi$. This is because the particles tend to move radially outward along the spiral magnetic field, as discussed above. This anisotropy becomes stronger in later time and for higher energy particles (see Section \ref{sec:conf}). Since this anisotropy appears in the particle simulations with all the MHD data sets, the grid spacing and resolutions are not the cause of the anisotropy.

\subsubsection{Diffusion in energy space}

  \begin{figure}
   \begin{center}
    \includegraphics[width=\linewidth]{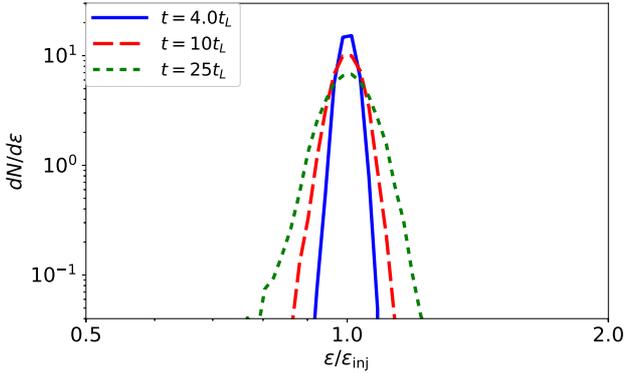}
    \caption{Energy distribution function at $t=4t_L$, $10t_L$, and $25t_L$ in fluid flame for $\lambda_{\rm ini}=4$. The distribution function diffuses in the energy space. }
    \label{fig:gamdist}
   \end{center}
  \end{figure}

  \begin{figure}
   \begin{center}
    \includegraphics[width=\linewidth]{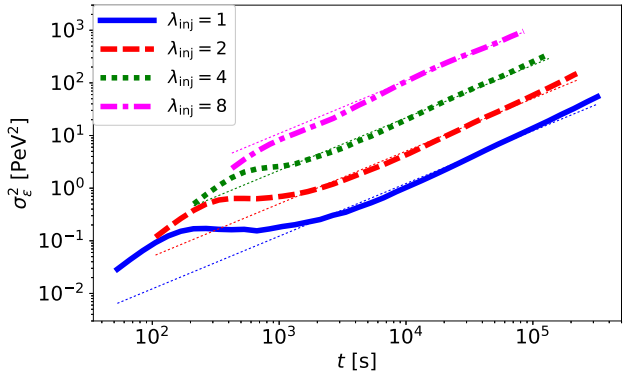}
    \caption{Time evolution of the variance of the particle energy, $\sigma_\epsilon^2$ for $\lambda_{\rm ini}=1$, (blue-slid line), 2  (red-dashed line), 4 (green-dotted line), and 8 (magenta-dot-dashed line). We can see that $\sigma_\epsilon^2\propto t$ for $t\gtrsim t_{\rm int}$.}
    \label{fig:siggam}
   \end{center}
  \end{figure}

  \begin{figure}
   \begin{center}
    \includegraphics[width=\linewidth]{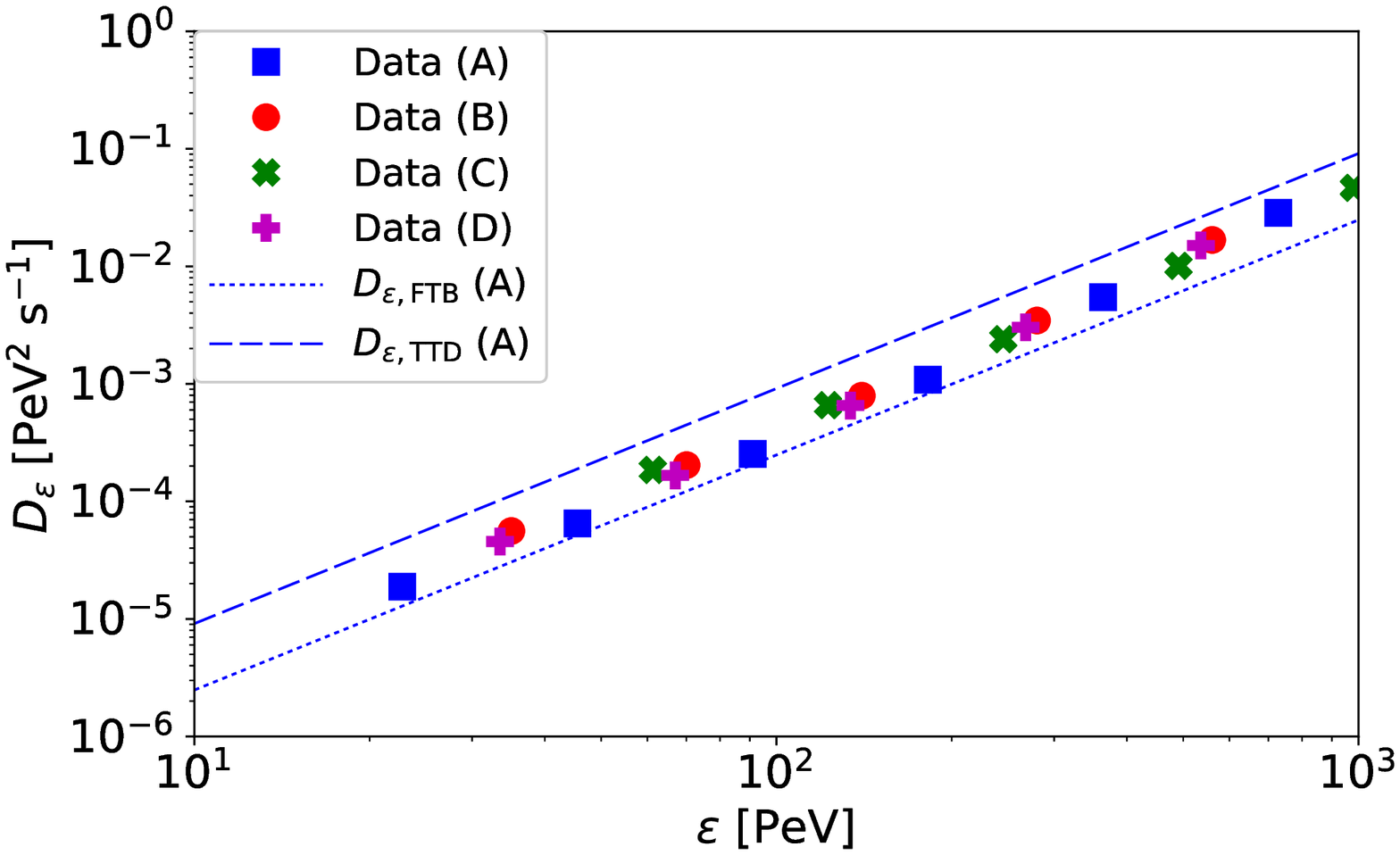}
    \includegraphics[width=\linewidth]{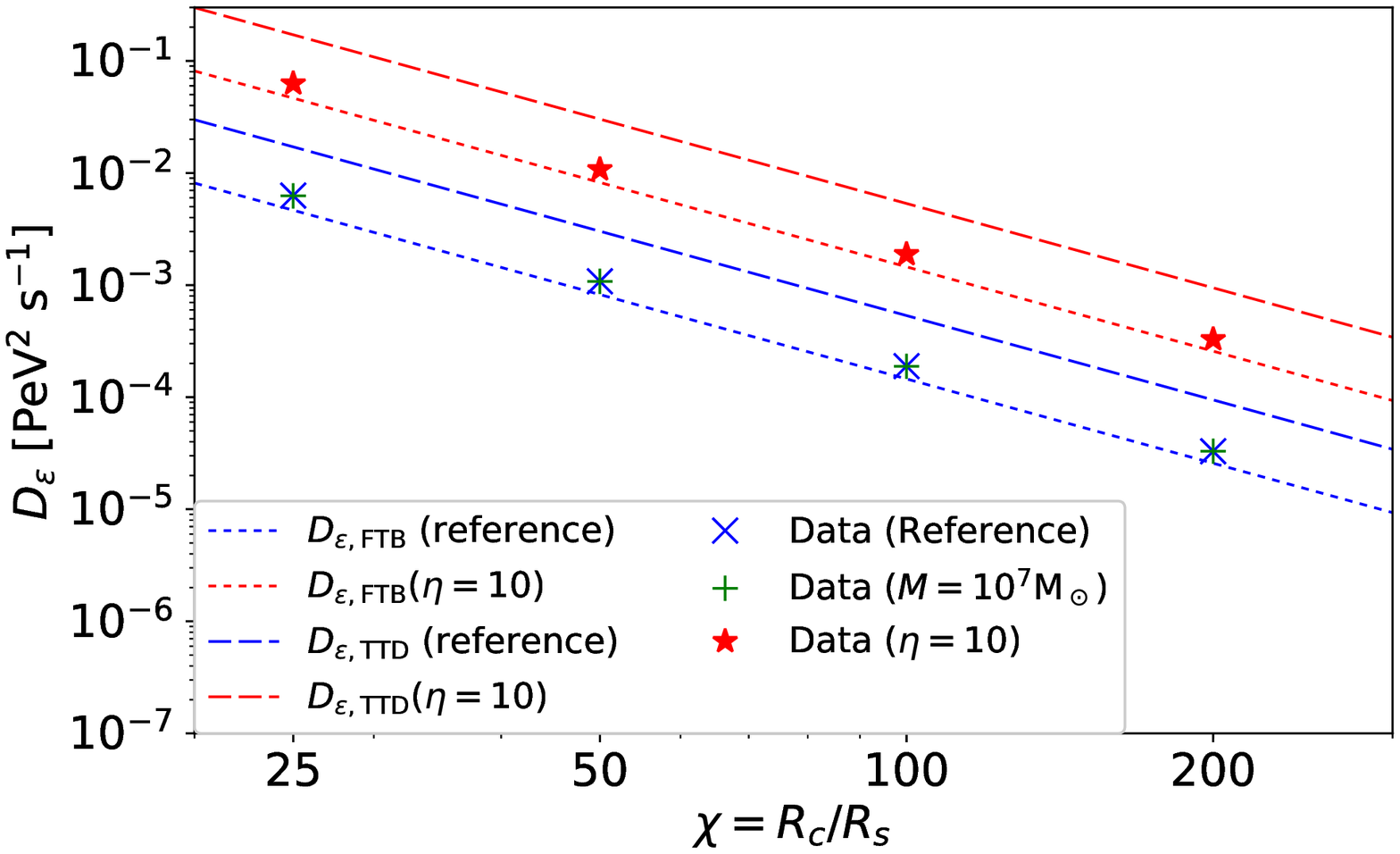}
    \caption{Parameter dependence of $D_\epsilon$. The upper panel shows the energy dependence of $D_\epsilon$ with various resolutions. The lower panel depicts the dependence on $\chi$ with various $\eta$ and $M$. We use $\lambda_{\rm ini}=4$ and the MHD data set of run A. We can see $D_{\epsilon,\rm FTB}$ is consistent with the simulation results within a factor of 2. $D_{\epsilon,\rm TTD}$ is also consistent within a factor of 3. }
    \label{fig:Dep}
   \end{center}
  \end{figure}

We examine evolution of the energy distribution function in the fluid frame. The time evolution of the energy distribution for $\lambda_{\rm ini}=4$ is shown in Figure \ref{fig:gamdist}. We can see that the width of the energy distribution increases with time. This motivates us to consider the diffusion equation in the energy space.

In general, the transport equation, including the diffusion and advection terms in both configuration and momentum spaces, describes the evolution of the distribution function for the particles with isotropic distribution in the fluid rest frame  \citep[e.g.][]{Ski75a,SMP07a}. When the terms for configuration space and the advection term in momentum space are negligible, the transport equation may be simplified to the diffusion equation only in momentum space \citep[e.g.,][]{sp08}: 
\begin{equation}
  \delf{f}{t}= {1\over p^2}\delp\left( p^2 D_p \delf{f}{p} \right).
\end{equation}
Since the anisotropy in our system is not very strong, we apply this equation to our system. We focus on the ultrarelativistic regime, so the particle energy is approximated to be $\epsilon\approx pc$. Using the differential number density, $N_\epsilon=N_p/c=4\pi p^2 f/c$, we can write the evolution of $N_\epsilon$ by the advection-diffusion equation in energy space: 
\begin{equation}
 \delf{N_\epsilon}{t} = \delep\left(D_\epsilon \delf{N_\epsilon}{\epsilon}\right)- \delep\left({2D_\epsilon\over \epsilon}N_\epsilon\right).\label{eq:dNpdt}
\end{equation}
In our simulation, most of the particles are confined in a narrow energy range of $\epsilon\sim \epsilon_{\rm ini}$. Thus, we approximate that $D_\epsilon\approx D_{\epsilon_{\rm ini}}$ and $2D_\epsilon/\epsilon \approx v_{\epsilon_{\rm ini}}$ are constant. Then, after some algebra, we obtain the time evolution of the mean and variance of the energy distribution: 
\begin{equation}
 \mu_\epsilon =  \frac{1}{N_{\rm actv}}\int N_\epsilon \epsilon d\epsilon \approx \epsilon_{\rm ini} + v_{\epsilon_{\rm ini}} t \label{eq:muep00}
\end{equation}
\begin{equation}
\sigma_\epsilon^2 = \frac{1}{N_{\rm actv}}\int N_\epsilon \epsilon^2 d\epsilon-\mu_\epsilon^2 \approx 2D_{\epsilon_{\rm ini}} t\label{eq:sigep00}
\end{equation}
where $N_{\rm actv}$ is the number of the particles confined in the computational region (see Appendix \ref{sec:derivation} for derivation).

Figure \ref{fig:siggam} shows the time evolution of $\sigma^2_\epsilon$ for $\lambda_{\rm ini}=1$, 2, 4, and 8 from $t=0.1t_L$ to the end of the simulation. Initially, $\sigma^2_\epsilon$ rapidly increases with time for all the models. Due to  the turbulent velocity component whose amplitude is about 10 \% of the background velocity, the particles are not exactly in the rest frame of the fluid elements, which causes drift motions. This leads to the initial jump of $\sigma^2_\epsilon$. For $\lambda_{\rm ini}=1$, 2, and 4, $\sigma^2_\epsilon$ becomes almost constant at $\sigma^2_\epsilon\sim 10^{-4}\epsilon_{\rm ini}^2$ after $t\gtrsim 0.3t_L$, because the particles start to move with the local drift velocity. In the late time, the particles start interacting with the turbulence, and $\sigma^2_\epsilon\propto t$ is realized. For $\lambda_{\rm ini}=8$, $\sigma^2_\epsilon$ continues to increase and approaches $\sigma^2_\epsilon\propto t$, because the gyration period is comparable to the interaction timescale with the turbulence. From the results, we find that the interaction timescale can be expressed as $t_{\rm int}\sim1\times10^3-2\times10^3M_8\chi_{1.7}$ s, including the parameter dependence. This is about a factor of 4 shorter than the crossing time of the turbulent length, $L_{\rm tur}/c\sim 7\times10^3M_8\chi_{1.7}$ s, i.e., we can write $t_{\rm int}\sim L_{\rm tur}/(4c)$.

This result, $\sigma^2_\epsilon\propto t$ for $t>t_{\rm int}$, indicates that the particle acceleration in the MRI turbulence occurs through the diffusion in energy space.
In the sub-sonic turbulence including the MRI turbulence, the slow modes are expected to play an important role in particle scattering \citep{lyn+14}. 
We can consider two mechanisms that change the particle energy in such turbulence:  the Fermi-type B mechanism  \citep[FTB; see e.g.][]{LPQ12a} and the transit-time damping (TTD). In FTB, the particles stream along a curved magnetic field that has a velocity. Then, the particles gain or lose energy at the fluid frame after the magnetic field sufficiently change the direction (see Figure 1 of \cite{LPQ12a}). The mean velocity of the magnetic field is expected to be $V_{R,\rm tur}$ in our MHD simulation. Then, the energy change per ``collision'' is approximated to be $\Delta\epsilon\sim \epsilon V_{R,\rm tur}/c$. Using the interaction time with the turbulence, $t_{\rm int}\approx L_{\rm tur}/(4c)$, the diffusion coefficient in energy space can be estimated to be \citep[e.g.,][]{BE87a}
\begin{eqnarray}
 D_{\epsilon,\rm FTB} &\approx& \frac13 \frac{\Delta \epsilon^2}{t_{\rm int}}\sim \frac{4\epsilon^2}{3} \frac{c}{L_{\rm tur}} \left({V_{R,\rm tur}\over c}\right)^2 \label{eq:DepFTB}\\
 &\propto& \epsilon^2M^{-1}\chi^{-2}\propto \lambda_{\rm ini}^2 \chi^{-5/2}\eta.\nonumber
\end{eqnarray}
In the last equation, we write down the parameter dependence using $\epsilon\approx\epsilon_{\rm ini}\propto\lambda_{\rm ini}M^{1/2}\chi^{-1/4}\eta^{1/2}$. We can use the relation because  the energy of the particles does not change very much in each run of the particle simulations. 

TTD requires the resonant condition: $v_{\rm pha}\simeq v_\parallel$, where $v_{\rm pha}$ is the phase velocity of the slow mode and $v_\parallel$ is the particle velocity parallel to the magnetic field. For the relativistic particles in weak sub-sonic turbulence, the condition for TTD cannot be satisfied, because $v_\parallel\sim c$ is always much faster than $v_{\rm pha}\sim V_A$. However, in strong turbulence, the relativistic particles can interact with the slow mode because non-linear effects broaden the energy range of the resonant particles \citep{YL08a,lyn+14}. If TTD is effective, the mean energy change per collision is typically $\epsilon V_A/c$. Then, the diffusion coefficient in energy space can be estimated to be
\begin{eqnarray}
 D_{\epsilon,\rm TTD} &\sim&\frac{\epsilon^2}{3} \left({V_A\over c}\right)^2 t_{\rm int}^{-1} \propto \epsilon^2M^{-1}\chi^{-2}.\label{eq:DepTTD}
\end{eqnarray}
The parameter dependence of $D_{\epsilon,\rm TTD}$ is the same as that of $D_{\epsilon,\rm FTB}$, while the normalization of $D_{\epsilon,\rm TTD}$ is higher than that of $D_{\epsilon,\rm FTB}$.

We calculate $D_\epsilon$ with various values of $\lambda_{\rm ini}=$ (0.5, 1, 2, 4, 8, 16), $M=$ ($10^7~\msun,~10^8~\msun$), $\chi=$ (30, 50, 100, 200), and $\eta=$ (1, 10), and show the resulting $D_\epsilon$ in Figure \ref{fig:Dep}. We combine the simulation results with various  $\epsilon_{\rm ini}$  to discuss the energy dependence of $D_\epsilon$. For the calculations with $\epsilon_{\rm ini}\gtrsim 10^3$ PeV, the particles escape from the computational region before the condition $\sigma^2_\epsilon\propto t$ is realized, so we only plot the results with $\epsilon_{\rm ini}<10^3$ PeV. The parameter dependence of $D_\epsilon$ is consistent with both of the simple estimates above: $D_\epsilon\propto \epsilon^2$ in the upper panel and $D_\epsilon\propto \chi^{-5/2}\eta$ for $\lambda_{\rm ini}=4$ in the lower panel. The normalization of the simple estimates are consistent with the simulation results within a factor of 3, while $D_{\epsilon,\rm FTB}$ matches better than $D_{\epsilon,\rm TTD}$. For the rest of the paper, we use $D_{\epsilon,\rm FTB}$ as a diffusion coefficient in energy space. The acceleration time is estimated to be 
\begin{equation}
t_{\rm acc}\approx{\epsilon^2\over 2D_{\epsilon,\rm FTB}}\sim \frac32 \left({c\over V_{R,\rm tur}}\right)^2t_{\rm int} \sim 1.7\times 10^7 M_{8} \chi_{1.7}^2 \rm s.\label{eq:tacc}
\end{equation}
This acceleration time is independent of energy.
Note that the fast modes have little influence on particle scattering in our simulation because they do not have enough power as discussed in Section \ref{sec:result-mhd}.

The values of $D_\epsilon$ can be estimated in two ways using either $\mu_\epsilon$ or $\sigma_\epsilon$, and these can be different when the particle distribution is anisotropic.
We evaluate the time evolution of $\mu_\epsilon$ via Equation (\ref{eq:muep00}), and confirm that the two methods are consistent with each other within a factor of 3 for $\lambda_{\rm ini}\lesssim 8$.
So far,  we have assumed that $D_\epsilon$ is constant, but our results indicate that $D_\epsilon\propto\epsilon^2$ is more realistic. 
For the case with $D_\epsilon\propto\epsilon^2$, we can derive the time evolution of $\mu_\epsilon$ and $\sigma_\epsilon^2$ without assuming that $D_\epsilon$ is constant. Then, as shown in Appendix,  the evolution of $\sigma_\epsilon^2$ is unchanged, while the increasing rate of $\mu_\epsilon$ is twice higher than that given by Equation (\ref{eq:muep00}):
\begin{equation}
\mu_\epsilon \approx  \epsilon_{\rm ini} + 2v_{\epsilon_{\rm ini}} t.\label{eq:muep01}
\end{equation}
We also estimate the values of $D_\epsilon$ based on the above Equation, and the results agree with those obtained by $\sigma_\epsilon^2$ within a factor of 2 for $\lambda_{\rm ini}\le8$ as shown in Appendix, implying the improvement compared to those based on Equation (\ref{eq:muep00}). For the models with $\lambda_{\rm ini}\gtrsim 8$, the anisotropy is large enough to affect the momentum diffusion, and the agreement becomes worse. However, this does not affect the discussion on the maximum energy in Section \ref{sec:discussion}, because in reality, high-energy particles escape from the system before they attain the energy corresponding to $\lambda_{\rm ini}\gtrsim 8$.


\subsubsection{Behaviour in configuration space}\label{sec:conf}

  \begin{figure}
   \begin{center}
    \includegraphics[width=\linewidth]{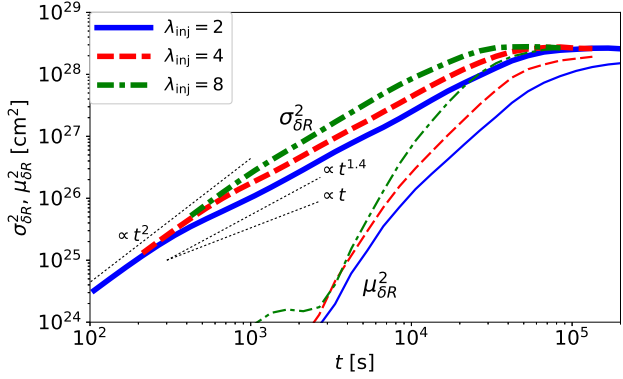}
    \caption{Time evolution of the mean and variance of the radial displacement. The thin and thick lines depict $\mu_{\delta R}^2$ and $\sigma^2_{\delta R}$  for $\lambda_{\rm ini}=2$ (solid), 4 (dashed), and 8 (dot-dashed), respectively. The dotted lines indicate the time dependence of $\sigma^2_{\delta R}$.  }
    \label{fig:sig_R}
   \end{center}
  \end{figure}

  \begin{figure}
   \begin{center}
    \includegraphics[width=\linewidth]{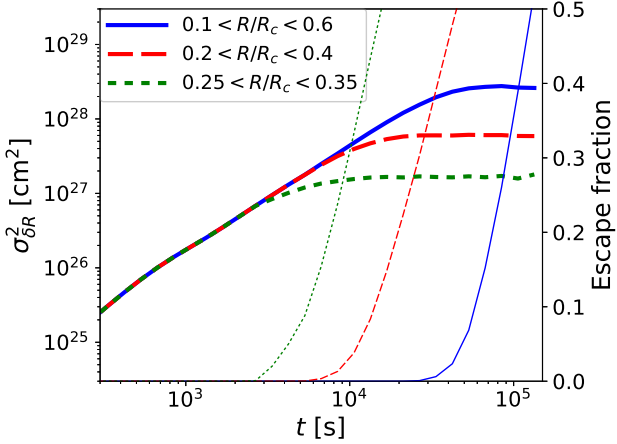}
    \caption{Time evolution of the variance of the radial displacement (thick lines) and the escape fraction (thin lines) for $\lambda_{\rm ini}=4$ with the various computational regions shown in the legend.}
    \label{fig:dx}
   \end{center}
  \end{figure}

  \begin{figure}
   \begin{center}
    \includegraphics[width=\linewidth]{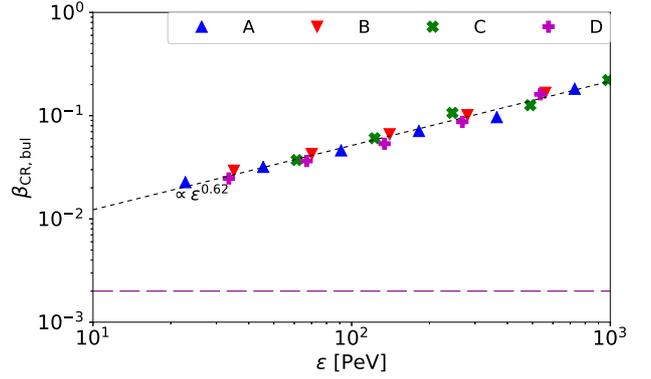}
    \includegraphics[width=\linewidth]{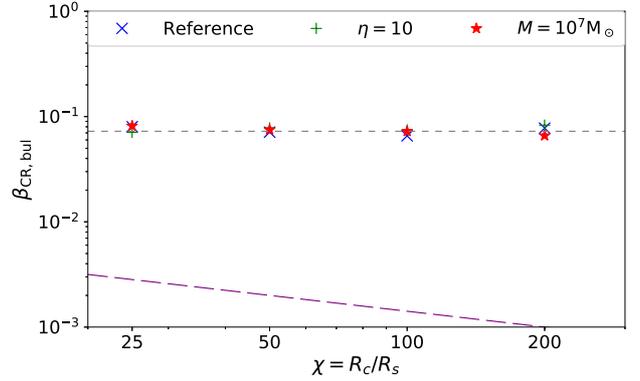}
    \caption{Parameter dependence of the radial bulk velocity of the CR particles, $\beta_{\rm CR,bul}$. The upper panel shows the energy dependence of $\beta_{\rm CR,bul}$.  The lower panel indicates dependence on the other parameters ($M$, $\chi$, $\eta$) with a fixed value of $\lambda_{\rm ini}=4$. The dotted lines are the fitting result represented by Equation (\ref{eq:vrave}). The dashed lines show the absolute values of the background fluid motion, $-V_r/c$. Note that the direction of the bulk CR motion is opposite to the background fluid motion.}
    \label{fig:vrave}
   \end{center}
  \end{figure}

  \begin{figure}
   \begin{center}
    \includegraphics[width=\linewidth]{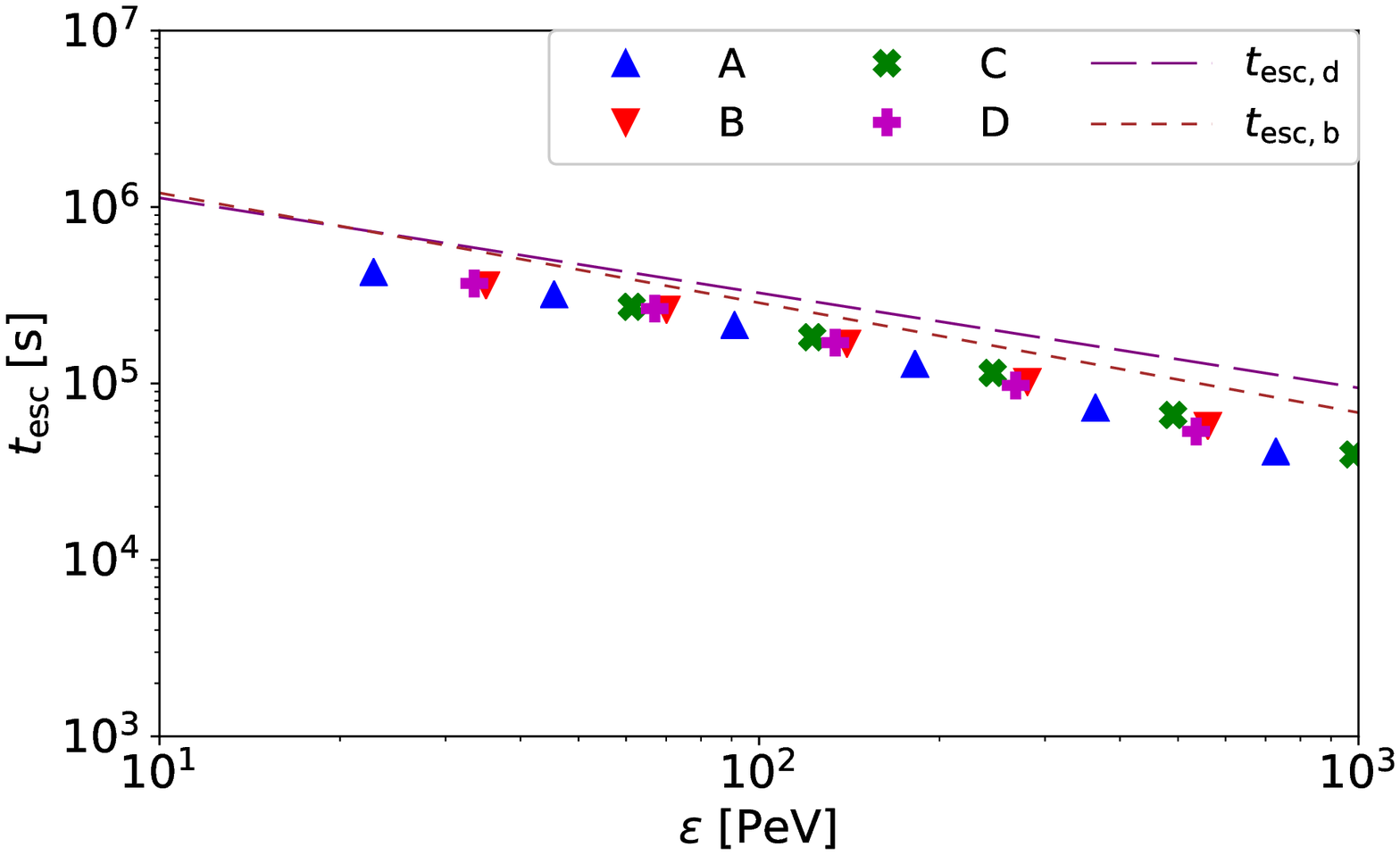}
    \includegraphics[width=\linewidth]{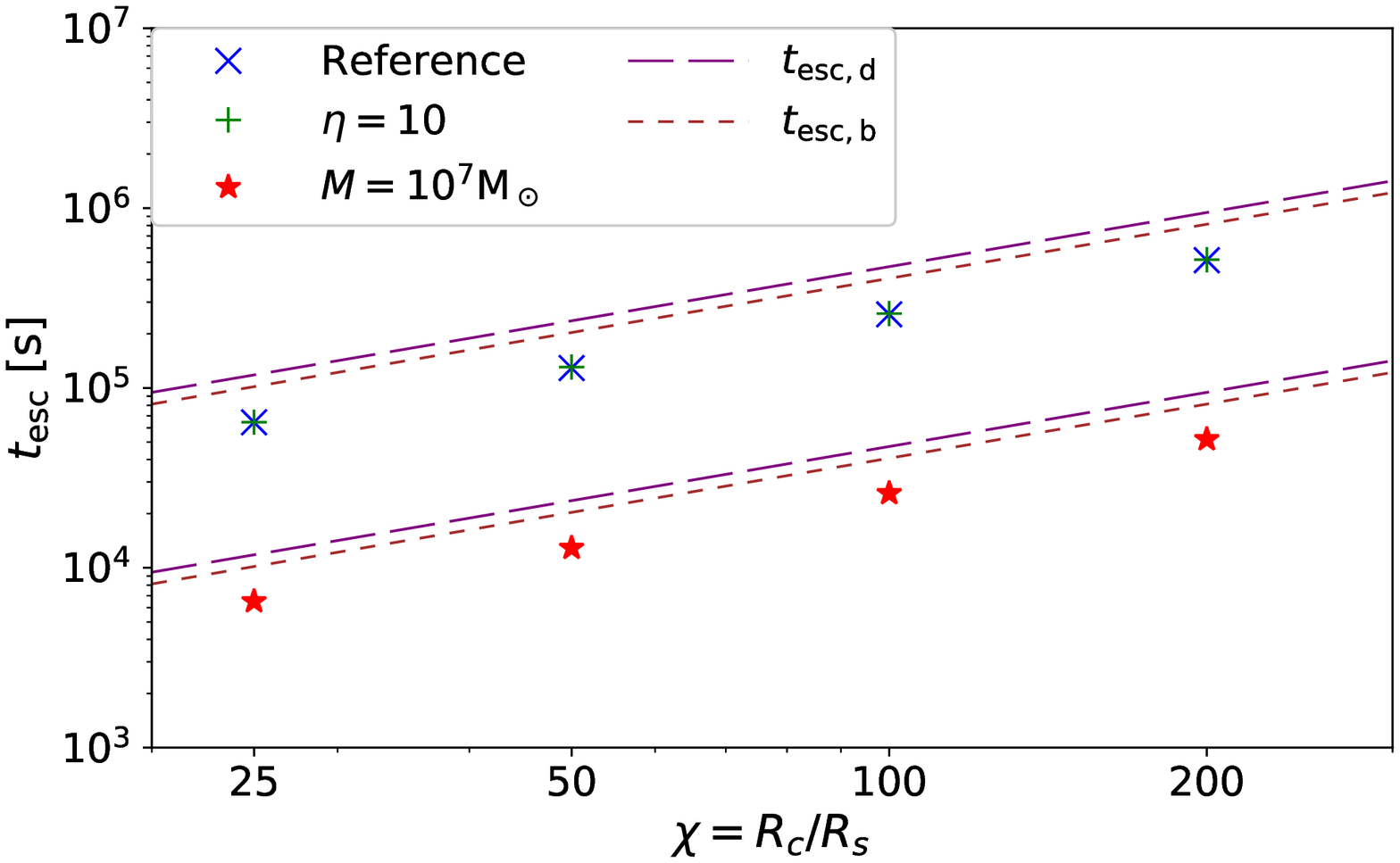}
    \caption{Parameter dependence of the escape timescale, $t_{\rm esc}$, at which half of the particles escape from the computational region. The upper panel shows the dependence on the particle energy, $\epsilon$. The lower panel indicates the dependence on the other parameters, $M$, $\chi$, and $\eta$ with a fixed value of $\lambda_{\rm ini}=4$. The dotted and dashed lines are the escape time given by Equations  (\ref{eq:tesc-adv}) and (\ref{eq:tesc-diff}), respectively.}
    \label{fig:tesc}
   \end{center}
  \end{figure}

  \begin{figure}
   \begin{center}
    \includegraphics[width=\linewidth]{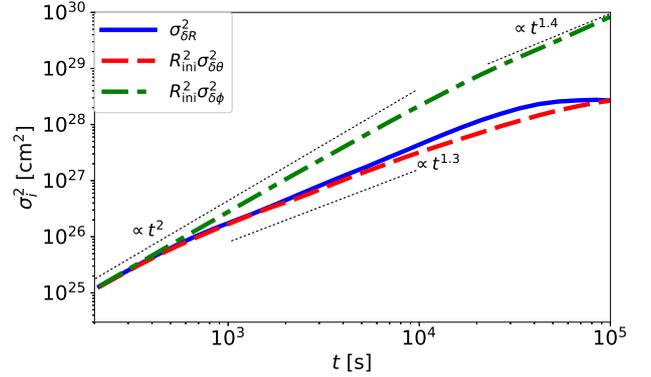}
    \caption{Time evolution of the variance of the displacement in configuration space, $\sigma^2_{\delta R}$ (solid line), $R_{\rm ini}^2\sigma^2_{\delta\theta}$ (dashed line), and $R_{\rm ini}^2\sigma^2_{\delta\phi}$ (dot-dashed line) for $\lambda_{\rm ini}=4$. The dotted lines indicate the time dependence.}
    \label{fig:sig_aniso}
   \end{center}
  \end{figure}

First, we discuss the displacement in $R$ direction, which is directly related to the escape process. We estimate time evolutions of the mean and the variance of the radial displacement:
\begin{eqnarray}
\mu_{\delta R}=\frac{1}{N_{\rm actv}} \sum_j \delta R_j,\\
\sigma^2_{\delta R}=\frac{1}{N_{\rm actv}} \sum_j \delta R_j^2 - \mu_{\delta R}^2,
\end{eqnarray}
where $\delta R_j=R_j-R_{\rm ini}$ is the radial displacement of each particle and $N_{\rm actv}$ is the number of the confined particles. Summation is performed over the particles confined in the computational region. $\mu_{\delta R}$ represents the bulk motion of CR particles, while $\sigma^2_{\delta R}$ expresses the diffusive motion.

 We show $\mu^2_{\delta R}$ and $\sigma^2_{\delta R}$ for the cases with various energies ($\lambda_{\rm ini}=2$, 4, and 8) in Figure \ref{fig:sig_R}. In the early phase, the diffusive motion is more efficient than the bulk CR motion. From the figure, we see that the stochastic behaviour of CRs in configuration space cannot be described as a usual diffusion. 
If the particles obey the usual diffusion, $\sigma^2_{\delta R}\propto t^2$ at the beginning, and $\sigma^2_{\delta R}\propto t$ after scattering timescale \cite[e.g.,][]{CLP02a,CM16a}.
In our simulation, $\sigma^2_{\delta R}$ initially increases with $t^2$. After about a half of the Larmor timescale, we see a transition to $\sigma^2_{\delta R}\propto t^{1.4}$, and finally, $\sigma^2_{\delta R}$ becomes flat at the time when the escape fraction becomes non-negligible. We analyze the data with various computational regions, and find that the behaviour is essentially the same; $\sigma^2_{\delta R}$ rapidly increases initially, and it is flattened when the escape becomes effective, as seen in Figure \ref{fig:dx}. This trend is similar for the cases with different parameter sets and other sets of the snapshot data of the MHD simulations. Thus, the radial variance is approximately written as
\begin{equation}
\sigma^2_{\delta R} \sim r_L^2\left({\zeta t\over t_L}\right)^\xi, \label{eq:sigmaR2}
\end{equation}
where $\xi\simeq 1.2-1.4$ and $\zeta\simeq 2-3$.
Since $1<\xi<2$, the particle behaviour in configuration space in the MRI turbulence is superdiffusive.
Note that $\sigma^2_{\delta R}$ starts to increase around $t\sim t_L/\zeta$, while $\sigma_\epsilon$ discussed in the previous subsection begins to grow at $t\sim t_{\rm int}$.

In later time, $\mu_{\delta R}^2$ rapidly increases with time, and becomes the dominant process for a higher energy run. This bulk motion originates from the anisotropic behaviour discussed in Section \ref{sec:orbits}. Using the distribution function, we can estimate the radial component of the bulk velocity of the CR particles:
\begin{equation}
 \beta_{\rm CR,bul}=\frac{c}{N_{\rm actv} \mu_{\epsilon}} \int  p_R \frac{dN_p}{dp_R} dp_R.
\end{equation}
For all the runs, $\beta_{\rm CR,bul}$ is not constant at the beginning. $\beta_{\rm CR,bul}$ starts to increase rapidly around $t\sim t_{\rm int}$, and for $t\gtrsim 5t_{\rm int}$, $\beta_{\rm CR,bul}$ becomes almost constant with time.
We estimate this constant values of $\beta_{\rm CR,bul}$ for various cases, and plot its energy dependence in the upper panel of Figure \ref{fig:vrave}. The higher energy runs have higher values of $\beta_{\rm CR,bul}$. The bottom panel shows the dependence on the other parameters for $\lambda_{\rm ini}=4$, where we see that $\beta_{\rm CR,bul}$ is independent of $M$, $\chi$, and $\eta$. According to our simulation results, the bulk radial velocity is represented as 
\begin{equation}
\beta_{\rm CR,bul}\simeq 2.9\times10^{-3}\epsilon_{\rm PeV}^{0.62}M_8^{-0.31}\chi_{1.7}^{0.16}\eta_0^{-0.31},\label{eq:vrave}
\end{equation}
where $\epsilon_{\rm PeV}=\epsilon/\rm PeV$.
This power-law representation fits the simulation data well as seen in Figure \ref{fig:vrave}. This bulk CR motion is much faster than the background MHD fluid, $V_R/c\sim 2\times10^{-3}\chi_{1.7}^{-1/2}$. Also, the bulk CR motion is outward, while the background fluid moves inward. Note that this energy dependence of the bulk CR motion appears due to our mono-energetic simulations. In reality, the CRs have an energy distribution, and convolution of CRs for all the energies will provide an energy-independent CR bulk velocity.

We estimate the escape time using $\sigma^2_{\delta R}$ and $\beta_{\rm CR,bul}$.
If the bulk CR motion is effective, the escape time can be estimated to be 
\begin{equation}
 t_{\rm esc,b}\sim \frac{R_{\rm esc}-R_{\rm ini}}{c\beta_{\rm CR,bul}}\propto  \lambda_{\rm ini}^{-0.62} M\chi  \propto \epsilon^{-0.62} M^{1.31}\chi^{0.84}\eta^{0.31}\label{eq:tesc-adv}
\end{equation}
On the other hand, for the diffusion-dominant cases, the particle starts to escape when $\sigma^2_{\delta R}\gtrsim (R_{\rm esc}-R_{\rm ini})^2$. Then, we can estimate the escape time to be
\begin{eqnarray}
t_{\rm esc,d}\sim \left({R_{\rm esc}-R_{\rm ini}\over r_L}\right)^{2/\xi} \frac{t_L}{\zeta} \propto \lambda_{\rm ini}^{\xi-2\over\xi} \chi M \nonumber\\
\propto \epsilon^{\xi-2\over \xi}M^{2+\xi\over 2\xi}\chi^{5\xi-2\over 4\xi}\eta^{2-\xi\over 2\xi}. \label{eq:tesc-diff}
\end{eqnarray}
We expect $t_{\rm esc,d}\propto \epsilon^{-0.43}-\epsilon^{-0.82}$ with $\xi\simeq 1.2-1.4$. For the demonstration purpose, we use $\xi=1.3$ and $\zeta=3$ for the rest of the paper, which leads to $t_{\rm esc,d}\propto \epsilon^{-0.54}$ for fixed values of $M$, $\chi$, and $\eta$ and $t_{\rm esc,d}\propto M\chi$ for a fixed value of $\lambda_{\rm ini}$. Note that the dependences on $M$, $\chi$, and $\eta$ for a fixed value of $\epsilon$ differ from those for a fixed $\lambda_{\rm ini}$.

We calculate a typical escape time, $t_{\rm esc}$, at which half of the particles escape from the system, and show it as a function of the particle energy in the upper panel of Figure \ref{fig:tesc}\footnote{The definition of $t_{\rm esc}$ is arbitrary. If we define $t_{\rm esc}$ as the time when the number of the particles inside the computational region becomes $e^{-1}$ of the initial number, $t_{\rm esc}$ can be longer than those in Figure \ref{fig:tesc}. The difference between these two definition is within a factor of 1.4}. The higher-energy particles escape from the system faster than the lower ones. 
The dotted and dashed lines show the escape time given by Equation (\ref{eq:tesc-adv}) and (\ref{eq:tesc-diff}), respectively, which roughly fit the simulation results for $\epsilon \lesssim 10^3$ PeV. Interestingly, these two estimates give similar values.
The slope becomes steeper for higher-energy particles at $\epsilon\gtrsim 10^3$ PeV, because these particles escape before interacting with the turbulent field. In the lower panel of Figure \ref{fig:tesc}, we calculate $t_{\rm esc}$ with various parameter sets, and find that the parameter dependence of $t_{\rm esc}$ is $t_{\rm esc}\propto M \chi$ for a fixed value of $\lambda_{\rm ini}$, which is also consistent with both of the estimates. 
Note that this escape time is longer than the light crossing time, $0.3R_c/c\sim 1.5\times10^4M_8\chi_{1.7}$ s as long as $\epsilon\lesssim 10^3$ PeV. Thus, the escape process is not ballistic.  Note also that our results indicate that the advection by MHD fluid motion, either turbulent motion or outflow, is not a dominant process of the particle transport. The advection time by MHD fluid is represented as $t_{\rm MHD}\approx R/V_{\rm MHD}$, where $V_{\rm MHD}$ is the turbulent velocity or the outflow velocity. The parameter dependence of the advection time by the MHD fluid is $t_{\rm MHD}\propto M\chi^{3/2}$, which is inconsistent with our results.

Next, we discuss the anisotropic behaviour in configuration space. The variances of the polar and azimuthal displacements are estimated to be 
\begin{eqnarray}
\sigma_{\delta\theta}^2=\frac{1}{N_{\rm actv}} \sum_j \delta\theta_j^2 - \mu_{\delta\theta}^2,\\
\sigma_{\delta\phi}^2=\frac{1}{N_{\rm actv}} \sum_j \delta\phi_j^2 - \mu_{\delta\phi}^2,
\end{eqnarray}
where $\delta \theta_j=\theta_j-\pi/2$ and $\delta \phi_j=\phi_j-\phi_{j0}$ are the displacements in $\theta$ and $\phi$ directions, respectively ($\phi_{j0}$ represents the initial azimuthal position), $\mu_{\delta\theta}$ and $\mu_{\delta\phi}$ are the means of the displacements.
Figure \ref{fig:sig_aniso} plots $\sigma^2_{\delta R}$, $R_{\rm ini}^2\sigma^2_{\delta\theta}$, and $R_{\rm ini}^2\sigma^2_{\delta\phi}$ for $\lambda_{\rm ini}=4$.
Initially, all lines increase with $t^2$, and later, they changes the increasing rates to $t^{1.3}-t^{1.4}$. We can regard the break time as the mean free time for each direction. The mean free time for the azimuthal direction is around $10^4$ s, which is longer than those for the radial and polar directions ($\sim t_L/\zeta\simeq 7\times10^{2}$ s).  This means that the particles stream to $\phi$ direction without strongly being scattered for a longer time. Since the magnetic field is directed to the azimuthal direction, this result also indicates that particles tend to stream parallel to the magnetic field.  The mean free time in $\phi$ direction is the same order as the time when $\beta_{\rm CR,bul}$ becomes constant, $\sim 5t_{\rm int}$.  Since the particles escape to radial direction, $\sigma^2_{\delta\theta}$ and $\sigma^2_{\delta\phi}$ continue to increase even after the non-negligible fraction of the particles escape from the system. We also estimate $\mu_{\delta\theta}$ and $\mu_{\delta\phi}$, and find that the bulk motions in $\theta$ and $\phi$ directions are sub-dominant for the parameter space we investigated here.

To discuss whether scattering between waves and particles occurs, we often use the first adiabatic invariance, $\mu_{\rm ad}=p_{\perp,*}^2/(2mB_*)$, where $p_{\perp,*}$ is the momentum perpendicular to the magnetic field and the subscript $*$ indicates the values at the fluid frame. In our simulation, since the magnetic field is turbulent in the scale of the Larmor radius, the adiabatic invariance is not conserved. Hence, we cannot use $\mu_{\rm ad}$ for measuring the mean free time.

\section{Discussion}\label{sec:discussion}

\subsection{Maximum energy of CRs}

The escape time is given by Equations (\ref{eq:tesc-adv}) or (\ref{eq:tesc-diff}), and the acceleration time is written in Equation (\ref{eq:tacc}). We here discuss implications of these results, and estimate the maximum achievable energy as an example. Since the acceleration time is longer than the escape time shown in Figures \ref{fig:tesc}, the expected maximum energy is lower than those assumed in our simulations. For the lower energy particles, $t_{\rm esc,d}$ is shorter than $t_{\rm esc,b}$. Equating the acceleration time and the diffusive escape time, we obtain the maximum energy:
\begin{eqnarray}
\epsilon_{\rm max}\approx \left(\frac{3\zeta}{16\pi}\frac{c^2}{V_{R,\rm tur}^2}\frac{L_{\rm tur}}{r_{L,i}}\right)^{\xi\over\xi-2}\left(\frac{R_{\rm esc}-R_{\rm ini}}{r_{L,i}}\right)^{2\over2-\xi}\epsilon_i \nonumber \\
\sim 0.07 M_8^{1/2}\chi_{1.7}^{-2.1}\eta_0^{1/2} \rm~PeV \label{eq:epmax}
\end{eqnarray}
where $r_{L,i}$ is the Larmor radius for the particles of $\epsilon=\epsilon_i$ and we use $\xi=1.3$ and $\zeta=3$ to obtain the value. A higher BH mass makes the system size larger, which helps to accelerate CRs to higher energy. The magnetic field is stronger for a higher $\eta$ or smaller $\chi$, leading to the higher $\epsilon_{\rm max}$. Note that even if we use $t_{\rm esc,b}$ instead of $t_{\rm esc,d}$, the estimate does not drastically change.

For $\chi=10$, which corresponds to $R_{\rm ini}=3R_s$, the maximum energy is $\epsilon_{\rm max}\simeq 2$ PeV. With this energy, we may expect production of high-energy neutrinos of $\sim 0.1$ PeV from accretion discs through inelastic hadronic collision and photomeson production \citep{kmt15}. Indeed, the astrophysical neutrinos of $0.01-10$ PeV are detected by IceCube  \citep{ice15,ice15a}, and LLAGNs are a good candidate of the neutrino source \citep{kmt15,KG16a}. On the other hand, the maximum energy for Galactic X-ray binaries of $10~\msun$ is at most a few TeVs according to Equation (\ref{eq:epmax}). This means that hot accretion flows in Galactic X-ray binaries cannot produce neutrinos detected by IceCube through the turbulent acceleration mechanism.

Our results imply that the hot accretion flow cannot accelerate ultrahigh-energy CRs (UHECRs). A higher $\eta$, i.e., higher mass accretion rate, results in forming a standard thin disc \citep{ss73,OM11a}, where the Coulomb loss prevents the particles from being accelerated \citep{ktt14,kmt15}. This condition gives $\eta\lesssim 10$ (see Section \ref{sec:setup-particle}). The initial radius should be larger than a few $R_s$ for the CRs to escape from the system. Otherwise, the CRs should fall to the BH because the accretion flow is supersonic \citep[e.g.,][]{ACG96a,NKM97a,ktt14}. The maximum mass of a SMBH is expected to be $\sim 10^{10}~\msun$ \citep{Net03a,JIL15a,II17a}. Even with the extreme parameter set ($M=10^{10}~\msun$, $\chi=10$, $\eta=10$), the maximum CR energy is estimated to be $\sim 70$ PeV, which is too low to be the source of UHECRs \citep[see e.g.,][for a review]{KO11a}. They should be produced by other sites or sources, such as radio galaxies \citep[e.g.,][]{Tak90a,MDT12a,KMZ18a} or gamma-ray bursts \citep[e.g.,][]{Wax95a,MIN08a,ZMK18a}.

\subsection{Comparison to the shearing box simulations}

\citet{KTS16a} performed MHD simulations with the shearing box approximation, and investigated the behaviour of CRs using the snapshot data of the shearing box MHD simulations. In the shearing box simulations, the particle acceleration is described by a diffusion phenomenon in energy space. The troidal magnetic field dominates over the poloidal field, so the particles tend to move to the azimuthal direction. 

These features are common with the global simulations. However, we find a few discrepancies between the global and the shearing box simulations. The first point is the importance of the shear acceleration  \citep[see e.g.,][]{BK81a,Kat91a,SBK99a}. In the shearing box simulations, the shear acceleration is effective for higher energy particles. However, the shear acceleration is not observed in the global simulations. This is because the calculation region is finite for the global simulations. The shearing box approximation creates an unrealistic shear velocity due to the periodic boundary condition. With the global simulation data, the high-energy particles escape from the system before the acceleration by the shear. Therefore, the shear acceleration is not effective in the realistic accretion flows. 

Another is the bulk outward motion discussed in Section \ref{sec:orbits} and \ref{sec:conf}. In the global simulations, the magnetic fields have spiral structure owing to the differentially rotating accretion flow. Since the magnetic field is stronger in the inner region, the magnetic mirror force prevents the particles from moving inward. Therefore, the particles tend to move outward when they stream along the spiral magnetic field. By contrast, the shearing boxes do not distinguish radially inward or outward. Although the local simulations can create spiral magnetic fields, there is no magnetic mirror force because the magnetic field strength should be the same in both directions. Thus, they cannot produce the bulk outward motion. Note that the quantitative physical understanding of the energy dependence of $\beta_{\rm CR,bul}$ remains as future work.

The other is the appearance of the superdiffusion. In the shearing box simulations, the spacial diffusion is described by the anisotropic Bohm diffusion, where the diffusion coefficient for the azimuthal direction is higher than those for the other two directions. On the other hand, the global simulations result in a superdiffusive transport. While the superdiffusive transports are observed in some of the previous simulations \citep{ZPV06a,XY13a,RII16a} and have been discussed in the literature of interplanetary turbulence \citep{PZ07a,PZ09a,SS11b}, their situations are different from our setup of the turbulent fields. Understanding the cause of the superdiffusion is left as future work.

\subsection{Future directions}

Our results indicate $D_\epsilon\propto \epsilon^2$ with $\lambda_{\rm ini}\simeq0.5-8$. This is different from the index of the turbulent power spectrum in the relevant scales ($m\gtrsim 100$).  This means that the resonant scattering is ineffective in our simulation. However, the numerical dissipation suppresses the turbulent power in these small scales. If the turbulence has sufficient power in the smaller scales, the resonant scattering might be important for the lower energy particles. Higher resolution calculations with a higher-order reconstruction scheme is necessary to investigate effects of the waves at the smaller scales. 

 In the vicinity of the BH, our Newtonian treatment is no longer valid, and effects of general relativity must be important.
Many general relativistic MHD (GRMHD) simulations are peformed \citep[e.g.,][]{Mck06a,TOK16a}. However, GRMHD calculations require more computational time than Newtonian ones, which makes it difficult to perform high resolution simulations. Also, in the GRMHD simulations, the velocity of the background fluid is close to the speed of light. This situation disallows our use of the snapshot data, so we need to perform particle simulations with a dynamically evolving turbulence.

Although we setup the initial condition for MHD simulations with the zero net vertical magnetic flux, there can be a strong large-scale magnetic field \citep{NIA03a,TNM11a,MTB12a}. This type of magnetic field configuration is expected to be related to the relativistic jet production. The non-thermal particles can play an important role on the mass loading processes to the jet \citep{tt12,ktt14}, which would be investigated by the test-particle simulations with GRMHD simulation data \citep{BRP18a}.

Hot accretion flows are expected to be composed of collisionless plasma \citep{tk85,qg99,ktt14}, where the non-ideal MHD processes can be important. Resistivity creates an electric field parallel to a magnetic field in a dissipation scale, which might affect the energy of CRs. The features of the MRI turbulence can be affected by the anisotropic pressure \citep{QDH02a,SHQ06a,KSS14a,HH17a} and/or anisotropic heat transfer \citep{RTQ15a,FCG17a}. Magnetic reconnections are expected to have some influence on the MRI turbulence, since they are important dissipation processes  \citep[e.g.,][]{SI01a}. The reconnections also provide seed CRs \citep{KGL11a,KGL12a,Hos12a,BOP17a}, which are necessary for the turbulent stochastic acceleration to work as CR sources.

\section{Summary}\label{sec:summary}

We have investigated behaviours of high-energy relativistic particles in MRI turbulence in accretion flows. We have generated the turbulent fields via MHD simulations to model accretion flows around a black hole, calculated orbits of test-particles by using the snapshot MHD data, and investigated the particle acceleration and escape processes. 

We have run four MHD simulations with the same initial condition of an equilibrium torus with different resolutions and grid spacing. For all the simulations, the MRI grows in a few rotation time of the initial torus, and a quasi-steady state is achieved after several rotation time. Due to the shear motion, $B_\phi$ is stronger than the other components. The turbulent velocity is sub-sonic and sub-Alfvenic, while it is faster than the radial advection velocity. The mass-density and magnetic-energy fluctuations are weakly anti-correlated, which means that the slow and Alfven modes have important influences on the turbulence. The power spectra of the magnetic fields are anisotropic: $mP_m\propto m$ for $B_r$ and $B_\theta$, while $mP_m\propto m^{1/2}$ for $B_\phi$, although they have a similar peak around $m\sim10-20$. These features of the turbulence are common for all the runs, although the lowest-resolution run has weaker turbulent fields because of the lack of resolution.

Using the snapshot data of the MHD simulations, we have calculated a number of orbits of test-particles. We have found that the particles tend to escape through the outer radial boundary, rather than through the vertical boundary or falling to the BH. The particles have anisotropic momentum distribution due to the magnetic field configuration generated by the shearing accretion flow. The evolution of the distribution function in energy space is described by a diffusion phenomenon. The diffusion coefficient is described by Equation (\ref{eq:DepFTB}), which implies that large-scale waves have a dominant role for changing the particle energy for the energy range we investigated. The evolution in configuration space is not simply written by the diffusion. The particles spread faster than the usual diffusion (superdiffusion). The variance of the radial displacement is approximated by Equation (\ref{eq:sigmaR2}). Also, the anisotropic momentum distribution creates the bulk outward motion of the CR particles. This bulk motion is faster for higher energy particles as shown in Figure \ref{fig:vrave}. For higher energy particles, the bulk CR motion is more effective than the diffusive motion, and vice versa for lower energy particles. Physical interpretations of the emergence of the superdiffusion and the energy dependence of the bulk CR motion are left as future work. To obtain a solid conclusion, MHD simulations with a higher-order scheme and an order of magnitude higher resolution is essential.











\section*{Acknowledgements}
S.S.K. thanks Matthew W. Kunz for useful comments. We are grateful to the anonymous referee for a careful reading of the manuscript and thoughtful comments. This work is supported by JSPS Oversea Research Fellowship, the IGC post-doctoral fellowship program (S.S.K.), Alfred P. Sloan Foundation, NSF Grant No. PHY-1620777 (K.M.), and JSPS KAKENHI Grant Numbers 16H05998 and 16K13786 (K.T.). Numerical computations were carried out on Cray XC30 and Cray XC50 at Center for Computational Astrophysics, National Astronomical Observatory of Japan.

\bibliographystyle{mnras}
\bibliography{ssk}

\appendix
\section{Derivation of Relation between $\sigma_\epsilon^2$ and $D_\epsilon$}\label{sec:derivation}

Under the approximation of $D_\epsilon\approx D_{\epsilon_{\rm ini}}$ and $2D_\epsilon/\epsilon\approx v_{\epsilon_{\rm ini}}$, Equation (\ref{eq:dNpdt}) is expressed as
\begin{equation}
 \delf{N_\epsilon}{t}=D_{\epsilon_{\rm ini}}{\partial^2 N_\epsilon\over \partial \epsilon^2} - v_{\epsilon_{\rm ini}}\delf{N_\epsilon}{\epsilon}.
\end{equation}
The mean of the momentum is written as $\mu_\epsilon = \int N_\epsilon \epsilon d\epsilon / N_{\rm actv} $. Its time derivative is 
\begin{eqnarray}
\df{\mu_\epsilon}{t} &=& {1\over N_{\rm actv}}\int \epsilon \delf{N_\epsilon}{t} d\epsilon \nonumber \\
&\approx& {1\over N_{\rm actv}}\int\left(\epsilon D_{\epsilon_{\rm ini}}{\partial^2 N_\epsilon\over \partial \epsilon^2} - \epsilon v_{\epsilon_{\rm ini}}\delf{N_\epsilon}{\epsilon}\right)d\epsilon \nonumber \\
&=& {1\over N_{\rm actv}}\int\left(-D_{\epsilon_{\rm ini}}{\partial N_\epsilon\over \partial \epsilon} + v_{\epsilon_{\rm ini}}N_\epsilon\right)d\epsilon \nonumber \\
 &=& v_{\epsilon_{\rm ini}},
\end{eqnarray}
where we use a partial integration and $N_\epsilon \to 0$ for $\epsilon\to \infty$ and $\epsilon\to 0$. Integrating both sides with $t$, we obtain
\begin{equation}
 \mu_\epsilon \approx \epsilon_{\rm ini} + v_{\epsilon_{\rm ini}}t \label{eq:muep00a} 
\end{equation}
A similar calculation gives us the variance $\sigma_\epsilon^2$. The variance of the momentum is written as $\sigma_\epsilon^2= \int N_\epsilon \epsilon^2 d\epsilon/N_{\rm actv}-\mu_\epsilon^2$. Its time derivative is
\begin{eqnarray}
\df{\sigma_\epsilon^2}{t} &=& {1\over N_{\rm actv}}\int \epsilon^2 \delf{N_\epsilon}{t} d\epsilon - 2 \mu_\epsilon\df{\mu_\epsilon}{t}  \nonumber \\
&\approx& {1\over N_{\rm actv}}\int\left(\epsilon^2D_{\epsilon_{\rm ini}}{\partial^2 N_\epsilon\over \partial \epsilon^2} - \epsilon^2 v_{\epsilon_{\rm ini}}\delf{N_\epsilon}{\epsilon}\right)d\epsilon - 2\mu_\epsilon v_{\epsilon_{\rm ini}} \nonumber \\
&\approx& {1\over N_{\rm actv}}\int\left(-2\epsilon D_{\epsilon_{\rm ini}}{\partial N_\epsilon\over \partial \epsilon} +2 \epsilon v_{\epsilon_{\rm ini}}{N_\epsilon}\right)d\epsilon - 2\mu_\epsilon v_{\epsilon_{\rm ini}} \nonumber \\
&=& {2 D_{\epsilon_{\rm ini}}}.
\end{eqnarray}
Hence, we obtain 
\begin{equation}
 \sigma_\epsilon^2 \approx 2 D_{\epsilon_{\rm ini}} t.\label{eq:sig0}
\end{equation}

For the special case of $D_\epsilon=D_{\epsilon_{\rm ini}} (\epsilon/\epsilon_{\rm ini})^2$, we can derive the time evolution of $\mu_\epsilon$ and $\sigma_\epsilon^2$ by similar algebra without assuming that $D_\epsilon$ is constant. Using Equation (\ref{eq:dNpdt}), the time derivative of $\mu_\epsilon$ is written by
\begin{eqnarray}
\df{\mu_\epsilon}{t} &=& {1\over N_{\rm actv}}\int \epsilon \delf{N_\epsilon}{t} d\epsilon \nonumber \\ 
&=& \int \left[\epsilon\delep \left(D_{\epsilon_{\rm ini}}\left(\frac{\epsilon}{\epsilon_{\rm ini}}\right)^2\delf{N_\epsilon}{\epsilon}\right)-\epsilon\delep\left(\frac{2D_{\epsilon_{\rm ini}}\epsilon}{\epsilon_{\rm ini}^2} N_\epsilon\right)\right]\nonumber\\
&=& \frac{4D_{\epsilon_{\rm ini}}}{\epsilon_{\rm ini}^2}\mu_\epsilon \nonumber\\
&=& 2v_{\epsilon_{\rm ini}} \mu_\epsilon,\label{eq:muep}
\end{eqnarray}
where $v_{\epsilon_{\rm ini}}=2D_{\epsilon_{\rm ini}}/\epsilon_{\rm ini}$. Then, we obtain
\begin{equation}
 \mu_\epsilon = \epsilon_{\rm ini}\exp\left(\frac{4D_{\epsilon_{\rm ini}}}{\epsilon_{\rm ini}^2}t\right)\approx \epsilon_{\rm ini}+2v_{\epsilon_{\rm ini}} t, \label{eq:muep01a}
\end{equation}
Hence, the increasing rate of $\mu_\epsilon$ is twice higher than that in Equation (\ref{eq:muep00a}).
The time derivative of the mean of $\epsilon^2$, $\langle \epsilon^2\rangle$, is
\begin{equation}
 \df{\langle \epsilon^2\rangle}{t} = {1\over N_{\rm actv}}\int \epsilon^2 \delf{N_\epsilon}{t} d\epsilon 
= 10\frac{D_{\epsilon_{\rm ini}}}{\epsilon_{\rm ini}^2}\langle \epsilon^2\rangle. 
\end{equation}
Integrating both sides, we get 
\begin{equation}
 \langle\epsilon^2\rangle = \epsilon_{\rm ini}^2\exp\left(\frac{10D_{\epsilon_{\rm ini}}}{\epsilon_{\rm ini}^2}t\right)\approx \epsilon_{\rm ini}^2+10D_{\epsilon_{\rm ini}}t.\label{eq:muep2}
\end{equation}
Using Equations (\ref{eq:muep01a}) and (\ref{eq:muep2}), we obtain
\begin{equation}
 \sigma_\epsilon^2 = \epsilon_{\rm ini}^2\left[\exp\left(\frac{10D_{\epsilon_{\rm ini}}}{\epsilon_{\rm ini}^2}t\right) - \exp\left(\frac{8D_{\epsilon_{\rm ini}}}{\epsilon_{\rm ini}^2}t\right)\right]\approx 2D_{\epsilon_{\rm ini}}t.
\end{equation}
This $\sigma_\epsilon^2$ is the same as that in Equation (\ref{eq:sig0}).

\begin{figure}
   \begin{center}
    \includegraphics[width=\linewidth]{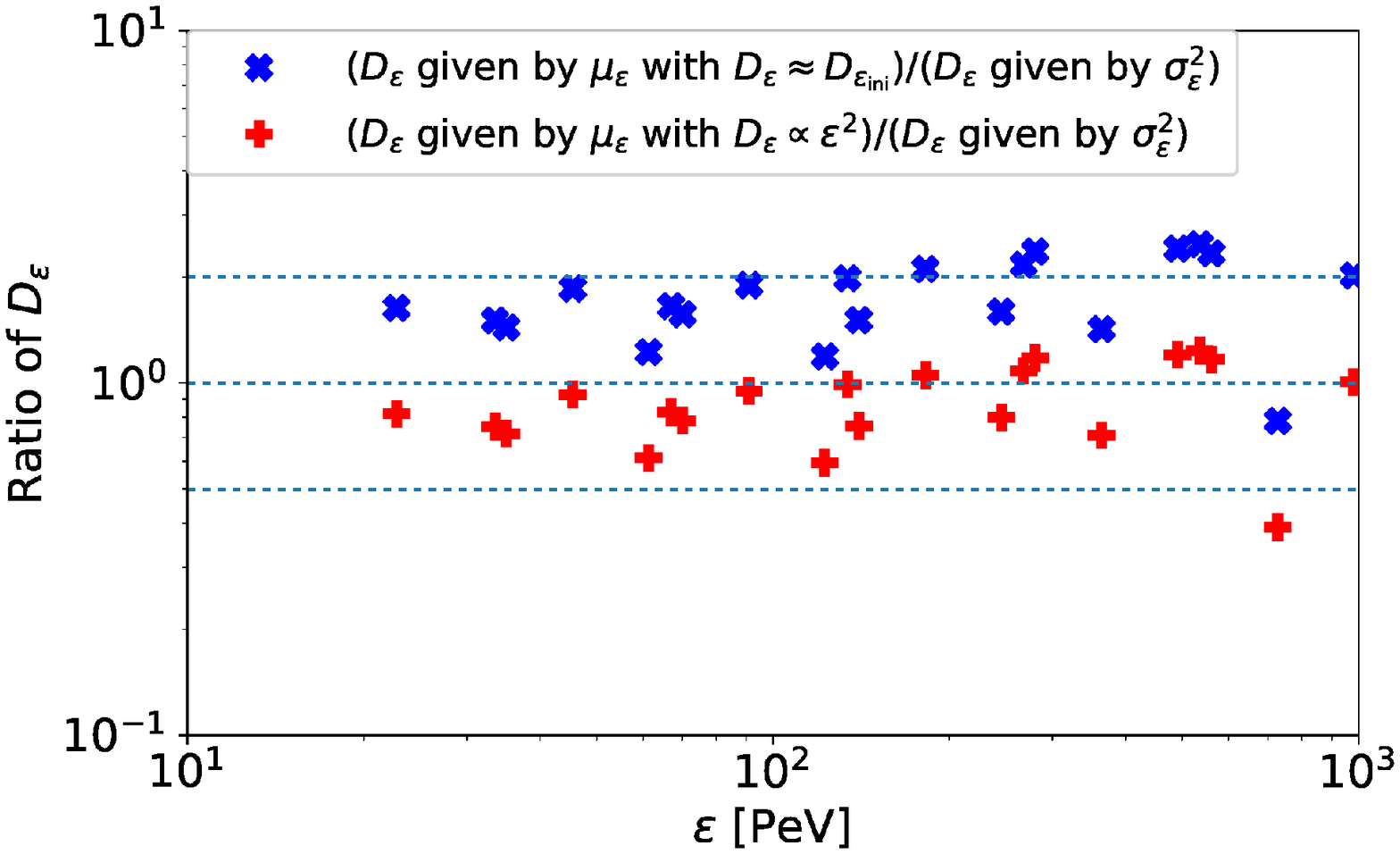}
    \caption{Ratio of $D_\epsilon$ obtained by Equations (\ref{eq:muep00a}) or (\ref{eq:muep01a}) to those given by $\sigma_\epsilon^2$. The blue-cross symbols are for the cases that $D_\epsilon$ is constant, while the red-plus symbols are for the cases with $D_\epsilon\propto \epsilon^2$.  The horizontal dotted lines show the ratio equal to 0.5, 1, and 2 for comparison.   }
    \label{fig:Dgam-Vgam}
   \end{center}
  \end{figure}

Figure \ref{fig:Dgam-Vgam} shows the ratio of $D_\epsilon$  obtained by $\mu_\epsilon$, using either Equation (\ref{eq:muep00a}) or Equation (\ref{eq:muep01a}), to those given by $\sigma_\epsilon^2$. We can see that the values are consistent within a factor of 3 for $\lambda_{\rm ini}\le 8$ (corresponding to $\epsilon\le3.6\times10^2$ PeV for run A) for the cases that $D_\epsilon\approx D_{\epsilon_{\rm ini}}$ is constant (blue cross symbols). The agreement is improved for the cases with $D_\epsilon\propto \epsilon^2$  for $\lambda_{\rm ini}\le 8$, where the values of $D_\epsilon$ agree each other within a factor of 2  (red plus symbols).

\label{lastpage}

\end{document}